\documentclass{article}
\usepackage{graphicx} 

\usepackage[dvipsnames]{xcolor}
\usepackage[normalem]{ulem}
\usepackage{parskip}

\usepackage{dirtytalk}
\usepackage[title]{appendix}
\usepackage[hyphens]{url}

\usepackage{authblk}
\usepackage{hyperref}
\usepackage[font=small,labelfont=bf]{caption}
\hypersetup{breaklinks=true}

\makeatletter
\renewcommand\AB@affilsepx{ \protect\Affilfont}
\makeatother

\newcommand*\samethanks[1][\value{footnote}]{\footnotemark[#1]}

\newcommand{\startappendix}{%
  \clearpage\appendix
  \setcounter{section}{0}%
  \renewcommand{\thesection}{\Alph{section}}%
}

\title{Risk Sources and Risk Management Measures in Support of Standards for General-Purpose AI Systems}

\author[1,2]{Rokas Gipiškis\thanks{Equal contribution. Correspondence: inquiries@aistandardslab.org}}
\author[1]{Ayrton San Joaquin\samethanks}

\author[1]{Ze Shen Chin\thanks{Authors ordered randomly}}
\author[1]{\authorcr Adrian Regenfuß\samethanks}
\author[1]{Ariel Gil\samethanks}
\author[1,3]{Koen Holtman\samethanks}

\affil[1]{AI Standards Lab}
\affil[2]{Vilnius University}
\affil[3]{Holtman Systems Research}

\begin{document}

\newcommand{\colorRS}{\textcolor{Bittersweet}{\textbf{Risk source}}}
\newcommand{\colorRMM}{\textcolor{Cerulean}{\textbf{Risk management measure}}}

\newcommand{\rs}[1]{\colorRS: \emph{#1}\newline}
\newcommand{\rmm}[1]{\colorRMM: \emph{#1}\newline}
\newcommand{\rmmEX}[1]{\colorRMM \textcolor{Plum}{\textbf{ (Experimental)}}: \emph{#1}\newline}

 \date{}
\maketitle

\begin{abstract}
   \noindent  There is an urgent need to identify both short and long-term risks from newly emerging types of Artificial Intelligence (AI), as well as available risk management measures. In response, and to support global efforts in regulating AI and writing safety standards, we compile an extensive catalog of risk sources and risk management measures for general-purpose AI (GPAI) systems, complete with descriptions and supporting examples where relevant. This work involves identifying technical, operational, and societal risks across model development, training, and deployment stages, as well as surveying established and experimental methods for managing these risks. To the best of our knowledge, this paper is the first of its kind to provide extensive documentation of both GPAI risk sources \emph{and} risk management measures that are descriptive, self-contained and neutral with respect to any existing regulatory framework. This work intends to help AI providers, standards experts, researchers, policymakers, and regulators in identifying and mitigating systemic risks from GPAI systems. For this reason, the catalog is released under a public domain license for ease of direct use by stakeholders in AI governance and standards. \footnotemark
   \footnotemark
\end{abstract}

\footnotetext[1]{All authors contributed to this work as part of the AI Standards Lab, where all research activities were conducted.}
\footnotetext[2]{\textbf{Public domain declaration}: This paper has been submitted under the Creative Commons CC BY 4.0 Attribution Deed \cite{cc4}. However, we place the entire text of Section \ref{sec:safety_engineering} and the entire risk catalog from Sections \ref{Model development} to \ref{Impacts} in the public domain, as described by the Creative Commons CC0 1.0 Universal Deed \cite{cc0}. This means that we explicitly permit any party to copy text from these sections into other documents, with or without attribution, and we also allow any party to modify the copied text for their specific needs. While our “risk sources” and “risk management measures” text encodes a significant part of the known state-of-the-art in GPAI risk analysis and risk management in the literature, we have written it to avoid text covered by third-party copyright.}

\newpage
\tableofcontents
\newpage

\section{Introduction}
In recent years, there has been great interest in using “GPAI models,” predictive or generative models trained on a large corpus of data and usually obtained from the open internet, as a new way of building AI systems. Interest in the models is present in academia, in industrial research, and in the markets, all pointing to an urgent need to manage the emerging risks in these models or systems which deploy them.

This paper contributes to developing the sub-field of GPAI safety engineering. We write self-contained pieces of text that can be directly inserted into technical safety standards and codes of practice. These codes of practice and standards are currently being written by policymakers to support various regulatory initiatives concerning GPAI. Specifically, our work is driven by the need to inform standards and codes of practice that are integral in the implementation of AI (and GPAI) safety engineering requirements outlined in the EU AI Act \cite{aiact}. We aim to answer the following questions:

\begin{enumerate}
    \item What are the different sources of risk\footnotemark, both in the short-term and long-term, arising from the development and deployment of GPAI models or systems?
    
    \item What methods, both experimental and state-of-the-art, are available to manage the systemic risks of GPAI systems at various points in the supply chain?
\end{enumerate}

\footnotetext{We define the terms “risk” and “risk source” in Section \ref{sec:risk_management_terms}. Some alternate definitions exist, such as those that define risks as possible unwanted events, and risk sources as the fundamental and structural causes of such events.}

The answers to these questions take the form of sections that list individual “risk sources” and “risk management measures.”  Generically, we refer to these as risk items, and the list we produce is the risk catalog. While the risk items are primarily meant for GPAIs, they can apply to some AI systems of different risk levels and those that are designed for a narrower scope. This catalog covers a broad range of the GPAI value chain but is not exhaustive. We envision this paper as a tool that intended readers can use in research and risk management for GPAI, as well as in writing codes and standards for this field.

\subsection{Intended Audience} 
Our intended audiences are researchers, policymakers, regulators, safety engineers, standards experts, and risk management experts. 

We do not claim that all risks are equally important to address (e.g., equally likely or severe); it is up to the relevant actors to perform risk analysis for their specific situations. 

To provide structure to the catalog, we defined broad risk categories and placed each risk item into what we believe is a reasonable category. Our categorization is not intended to serve as a taxonomy, but rather as an aid in searching for relevant risks for specific cases. We recognize that some risk sources or risk management measures may belong to multiple categories\footnote{E.g., model evaluations (covered in Section \ref{Evaluations}) may be performed during the model development stage (Section \ref{Model development}) and the post-deployment stage (Section \ref{Deployment}).} and that some of the categories overlap. We leave it to the reader to decide on a taxonomy that suits them best, although we propose a taxonomy as an example in Appendix \ref{sec:risk_taxonomy}.

\subsection{Related Work on GPAI Safety Engineering}
Our work most closely resembles the AI Risk Repository \cite{slattery2024ai}, which provides a comprehensive directory of AI-related risks. They propose two taxonomies - a “Causal" Taxonomy concerned with the causal origin of risks and “Domain" Taxonomy concerned with the specific domain the risk falls into. Similar work can also be found in AVID \cite{avidmlAVID}, an open-source database for submitting risks, and the Generative AI Misuse Taxonomy \cite{marchal2024generative}, which provides a taxonomy based on empirical analysis of misuse of current generative AI technology. For comparison, our \textit{risk sources} are analogous to the \textit{risks} from the above databases. 

Our work differs from the above primarily by a) including risk management measures surveyed from the literature, and b) in providing our material in a public domain license, designed for direct cut-and-paste use in standardization or other regulatory efforts. 

Our work is also similar to MITRE Atlas \cite{wymberry2024approach} in that they both contain risk sources (“tactics" and “techniques") and risk management measures (“mitigation"), but narrower in scope to exclusively focus on cybersecurity and AI security. Finally, OWASP AI Security and Privacy Guide \cite{OWASP_2023}, ENISA Multilayer Framework for Good Cybersecurity Practices in AI \cite{polemi2023multilayer}, and Google Secure AI Framework \cite{blogIntroducingGoogles} provide guidelines on good practices in cybersecurity and AI security but do not provide a broader catalog of risks. For further reading on similar frameworks, we refer to \cite{barrett2023ai, pispa2024comprehensive}.

The standards and guideline documents ISO/IEC 42001, ISO/IEC 23894, ISO/\hspace{0pt}IEC 27001, and ISO/IEC 27002 also include lists of risk sources and/or risk management measures relevant to GPAI safety engineering. In this paper, we have mostly refrained from describing risk management measures already covered in these standards.

\subsection{Paper Structure}
Section \ref{sec:terminology} describes the terms and format we use to describe risk items. Section \ref{sec:safety_engineering} explains how this paper contributes to risk management and ongoing regulatory efforts. Meanwhile, the risk catalog spans Sections \ref{Model development} through \ref{Impacts}. Section \ref{Model development} covers the risk sources and risk management measures related to the model development stage. Section \ref{Evaluations} catalogs evaluation-related material. Section \ref{Attacks} catalogs attacks on GPAIs and GPAI failure modes. Section \ref{Agency} focuses on agentic AI systems. Section \ref{Deployment} is dedicated to the model deployment stage. Cybersecurity-related material is presented in Section \ref{Cybersecurity}. The impacts of AI systems are described in Section \ref{Impacts}. Finally, in Section \ref{sec:disc}, we analyze the risk items and discuss the challenges that providers face for AI risk management. We include additional material in the Appendix. Appendix \ref{sec:gpai_tech} provides background on how GPAI is distinct from previous paradigms of AI, as well as an overview of how the GPAI value chain relates to regulatory concerns. We discuss risks in benchmarking and autonomous AI in Appendices \ref{sec:benchmarking} and \ref{sec:autonomous}, respectively. Finally, our proposed risk taxonomy is in Appendix \ref{sec:risk_taxonomy}.

\section{Terminology}
\label{sec:terminology}
We define several terms to contextualize the contents of this paper.  The terms used here were chosen to align closely with the contents of the EU AI Act \cite{aiact}. 

\begin{itemize}
    \item \textbf{General-Purpose AI (GPAI) Model} - An AI model that can perform a broad range of distinct tasks. The use of “broad" here can refer to both within and outside knowledge domains. For example, a biology model that can solve chemical synthesis equations and generate treatment plans is limited to the biological domain but is considered a GPAI model because of its broad use. Similarly, another GPAI model is a model that can translate text between languages and can converse with users beyond translation. Large Language Models (LLMs) are a prominent kind of GPAI Model.

    The term “GPAI model” is somewhat specific to the EU AI Act as a regulatory initiative, which only applies to  models put  on the market \cite{aiact}. For regulatory discussions in the UK and the US, the terms “frontier model” or “foundation model" are often used instead. 
    
    \item \textbf{GPAI System} - A system that includes at least one GPAI model as its component. One example is a wearable device that integrates a GPAI model to interpret signals from the external sensors. Another example is when the system is completely digital: the GPAI model can be a component that interacts with different API services. A concrete example of the latter is ChatGPT, which involves a GPAI model (initially GPT-3.5 \cite{brown2020language}, then GPT-4 \cite{achiam2023gpt}) interacting with a content filtering API so that the system serves as a chatbot. Another class of example is AutoGPT \cite{Significant_Gravitas_AutoGPT} or OpenAI o1 \cite{OpenAISystemCard}, which use multiple models or instances of a model in concert.

    \item \textbf{GPAI Provider} -  An organization that develops a GPAI model directly, or places a product containing a GPAI model on the market, is referred to as an upstream or downstream provider, respectively. Development refers to modifications made at any stage of the AI model lifecycle. Providers are primarily characterized by the fact that they are not the end-users of the GPAI model. 
    
    \item \textbf{Systemic Risk} - Risk that threatens aspects of society, such as “public health, safety, public security, fundamental rights" \cite{aiact}, and whose harms can spread rapidly “across the value chain" \cite{aiact}.\footnote{Note that systemic risk has alternative definitions in non EU AI Act contexts such as finance, that can be somewhat broader.} 

    In the classification scheme of the EU AI Act, GPAI models and systems trained with more than $10^{25}$ FLOPs are presumed to pose systemic risk \cite{aiact}, as well as models possessing “high-impact capabilities" and other considerations on its societal impact as further defined in the AI Act \cite{aiact}. Notably, this classification scheme is inclusive, and applies by having any of the multiple conditions trigger the classification of the GPAI as posing a systemic risk. There is some diversity between the EU AI Act and other regulatory initiatives, e.g., the White House Executive Order on AI \cite{whitehouseExecutiveOrder} uses different thresholds and classification schemes.
\end{itemize}

\subsection{Risk Management Terms}
\label{sec:risk_management_terms}
To describe the risk management process, we use the following technical terms:
\begin{enumerate}
    \item \textbf{Risk}: the combination of the probability of an occurrence of harm and the severity of that harm \cite{aiact}.
    
    \item \textbf{Harm}: a negative event or negative social development entailing value damage or loss to people, property, and the environment\cite{isoHARMS}.\footnote{Where this “Harm” definition is, in our opinion, broad enough to cover all the systemic risks considered by the GPAI provisions of the AI Act.}

    \item \textbf{Risk source}: an element which alone or in combination has the potential to give rise to risk \cite{isoRiskSource}.\footnote{Some other international standards use the words “hazard” and ``hazardous situations" to denote essentially the same concept. We have no strong opinion on the choice but we will use “risk source” in this contribution.}

    \item \textbf{Risk management measure}: a measure that is designed to lower risk, either when applied alone or in combination with other measures. This can include identification, mitigation, or prevention of risk sources or individual risks relevant to a given system or class of systems.
\end{enumerate}
	 	 	 	
From the definitions above, it follows that risk sources are causally upstream of harms. Risk sources span a wide range of phenomena: some are purely technical or physical, while in sociotechnical systems, the source can involve actions that are performed (or fail to be performed) by human users or stakeholders. In some cases, risk sources may also be inadequacies of risk management measures, where they describe ways in which risk management measures may not achieve their intended outcome.

\subsection{Formatting of Risk Items}

The sections below contain structured lists of items where each item describes either a “risk source” or a “risk management measure.” Note that some risk management measures can be sources of risks themselves.

Each item is generally formatted in the following way \footnotemark:
\begin{itemize}
    \item \textbf{Item type} (either \colorRMM \space or \colorRS): \emph{item title}

    \item Description: Descriptive text, designed to answer at least the following question: \textit{“What exactly is the nature of this risk source, or risk management measure?”} The descriptive text may also contain further information like examples and references to the literature. In general, the text is written in a style similar to that found in a dictionary or an introductory textbook.   
    
    The description text is explicitly designed to be descriptive without being prescriptive. The text describing a specific measure does not include any information that tells the reader if or when the measure should be used. While such prescriptive information is a crucial component for any actionable regulation, our work is formatted so that the \textit{“what is it?”} discussion can be clearly separated from the \textit{“when to use it?”} discussion.
\end{itemize}

\footnotetext{While our item “title” and “description” are usually inspired by one or more sources from the open literature and the open internet, we have been careful to respect copyright by avoiding the direct copying of significant pieces of text from these sources into the “title” and “description” parts of the items.}

\subsection{State-of-the-art Knowledge}
Formal safety regulations for various activities and industries, as created and enforced by governments, typically require that safety engineering processes (like those in Figure \ref{fig:risk_management}) are carried out while taking into account the applicable state-of-the-art. One example of such a requirement can be found in Article 8 of the EU AI Act \cite{aiact}. 

For a more comprehensive definition of “state-of-the-art,” we refer to the definition used by the European Commission in their standardization request to the European Standardisation Organisations to support EU policy on AI:

\emph{\say{State-of-art should be understood as a developed stage of technical capability at a given time as regards products, processes and services, based on the relevant consolidated findings of science, technology and experience and which is accepted as good practice in technology. The state of the art does not necessarily imply the latest scientific research still in an experimental stage or with insufficient technological maturity.}} \cite{standardsrequest} 

When it comes to the GPAI-related risk sources documented in this catalog, we believe that knowledge of these risk sources is to be treated as part of the “state-of-the-art” in GPAI safety engineering.

However, some of the included risk management measures are of a more “experimental” nature, which we denote as \colorRMM \textcolor{Plum}{\textbf{ (Experimental)}}. While the literature often shows that they have an effect, the authors of the measures are typically careful to point out that there is little mature knowledge about the exact drawbacks of these methods, or that they consider the methods described to still be in a research stage.

\section{Safety Engineering Process Steps Supported by This Paper}
\label{sec:safety_engineering}
Safety engineering is widely understood as an iterative process in which the evaluation of risks associated with a product drives the selection of risk management measures to ensure the safe use or release of the product.  This process may also involve modifying the product itself. The iterative process should be ongoing, even after placement on the market.

\begin{figure}[t]
    \centering
    \includegraphics[width=.8\textwidth]{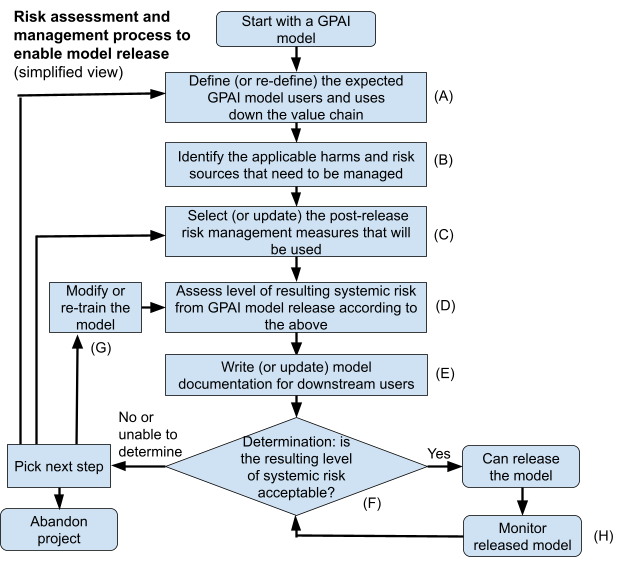}
    \caption{Iterative safety engineering process that can be used by GPAI model providers.  This paper has content that is specifically designed to support the engineering steps (B), (C), and (G). Note that we presented a simple flow of step (H), but it can vary depending on the monitoring strategy.}
    \label{fig:risk_management}
\end{figure}

Figure \ref{fig:risk_management} shows a sample graphical depiction of this process for the case of safety engineering leading to a positive or negative conclusion on releasing a GPAI model to its users or to the market, given a trained GPAI model. This depiction does not cover the initial steps to address the design and training phases of model development, nor does it detail any post-release steps.

The “risk source” content in this paper is specifically designed to support the engineering step (B), where “risk sources” need to be identified.

The key challenge in step (B) is to ensure that no potential risk source, or potential type of harm, is overlooked. Formal safety engineering processes always assume human fallibility. Therefore, absolute certainty can never be achieved in any of their steps. Safety engineering at the leading edge of technology is as much an art as it is a science, and practitioners typically use multiple tools to try to be as complete as possible. Hiring experienced specialists, who are able to map new situations to the knowledge of the general literature and of past incidents, is one such tool.  Another is to use “checklists”: to go over lists of risk sources, to significantly lower the probability that a risk is overlooked. The list of risk sources contained in this paper has been explicitly designed so that it can be used as a checklist with respect to GPAI risk sources. However, \textbf{our risk catalog is not exhaustive and any derivative of it should not be assumed as exhaustive}. Our risk items serve as a starting point to discover and map more risk sources.

The “risk management measure” content in this paper is specifically designed to support the engineering steps (C), (G), and (H). “risk management methods,” we wrote them from the perspective of either a member of the GPAI provider's \emph{model safety} engineering team or the \emph{system safety} engineering team. Readers may need to adjust the perspective depending on the context relevant to them.

For GPAI models and systems, step (F) is arguably the most difficult one: it is a step that has to combine both moral judgment and technical judgment. In the context of the EU AI Act, the “moral judgment” component should not be a unilateral decision by the safety engineering team itself, or by their management: instead it should involve an investigation into what society as a whole finds acceptable. In mature fields of safety engineering, as applied to long-existing products or systems, this investigation may be as easy as looking up certain design principles and threshold values in a technical standard that has been approved by the legislator: the standard may define that these principles and thresholds lead to morally acceptable outcomes. In the case of GPAI, and especially at the edge of innovation, no such easy shortcuts will be available. But legislators and standards writers can try to define tools and processes to make this easier. A further discussion of this complex topic is beyond the scope of this paper.

\section{Model Development}
\label{Model development}

This section catalogs the risk sources and risk management measures related to the model development stage. We categorize these into the following groups: data-related, training-related, and fine-tuning-related.

\subsection{Data-related}

\rmm{Documentation of data collection, annotation, maintenance practices}
Dataset collection, annotation, and maintenance processes can be documented in detail, including potential unintentional misuse scenarios and corresponding recommendations for data usage \cite{gebru2021datasheets, scheuerman2021datasets, jo2020lessons}. This contributes to transparency, ensures that inherent dataset limitations are known in advance, and helps in selecting the right datasets for intended use cases.

\rs{Difficulty filtering large web scrapes or large scale web datasets} A large scale “scraping” of web data for training datasets increases vulnerability to data poisoning, backdoor attacks, and the inclusion of inaccurate or toxic data \cite{foerderer2023should, birhane2021multimodal, cotroneo2024vulnerabilities}. With a large dataset, filtering out these quality issues is very difficult or trades off against significant data loss.

\rmm{Use of synthetic data} Synthetic data refers to data that is not collected from the real world. It is used to train AI models as an alternative to, or augmentation of, natural data. Effective use and generation of synthetic data allows for more oversight by the trainer on the training dataset because they have more control over its statistical properties. Synthetic data can help against dataset bias by having more samples from a particular distribution or minority group. It can also help in privacy by having more samples to mask sensitive data \cite{nikolenko2021synthetic}.

\rs{Lack of cross-organizational documentation} When sharing data between multiple organizations, documentation may be missing or inadequate, making it difficult for other organizations to understand it.

For example, a lack of metadata or a change in schema by a collaborating party can result in an unusable dataset and wasted data collection efforts, or it can lead to misunderstandings about the dataset's limitations, resulting in downstream risks related to its use \cite{sambasivan2021everyone}.

\rs{Manipulation of data by non-domain experts} Manipulating data (e.g., training data) carries a set of assumptions on how the data should appear and be used by those performing the manipulation. Common manipulations applied on data in the context of AI models include defining the ground truth label and merging different data formats or sources. People who have little or no expertise in the domain of the data performing such manipulations may render the data unusable or harmful to the development of the AI system \cite{sambasivan2021everyone}.

\rs{Insufficient quality control in data collection process} A lack of standardized methods and sufficient infrastructure, including the absence of quality control processes for collecting data, especially for high-stakes domains and benchmarks, can affect the quality and type of the data collected \cite{sambasivan2021everyone, jacovi2023stop}. This may include risks of dataset poisoning, inadvertent copyright violation, and test set leakages which invalidate performance metrics. 

\subsection{Training-related}

\rmmEX{Cost-inducing training of AI models specifically for malicious use} AI models can be designed to make further post-training modifications (e.g., fine-tuning) too costly for malicious uses while preserving normal adaptability for non-malicious uses \cite{henderson2023self, deng2024sophon}.

\rmm{Restrict web access during AI training} Developers can restrict or disable AI systems' internet access during training. For example, developers can restrict web access to read-only (e.g., by disabling write-access through HTTP POST requests and access to web forms) or limit the access of the AI system to a local network \cite{barrett2022actionable}.

This can prevent an AI system from overloading third-party web services by making too many requests, or posting inadequate or harmful content to the internet before being trained or fine-tuned not to produce harmful outputs. 

\rs{Adversarial examples} Adversarial examples \cite{szegedy2013intriguing, goodfellow2014explaining} refer to data that are designed to fool an AI model by inducing unintended behavior. They do this by exploiting spurious correlations learned by the model. They are part of inference-time attacks, where the examples are test examples. They generalize to different model architectures and models trained on different training sets.

Unintended behavior can range from incorrect predictions with respect to the ground-truth prediction to outputs that are generally considered undesirable (e.g., toxic or harmful). 

For example, when an autonomous vehicle’s sensor sees a stop sign with an adversarial sticker, the vehicle’s AI system may misclassify the stop sign as an indicator for the vehicle to accelerate \cite{eykholt2018robust}.

\rmm{Adversarial training} Adversarial training \cite{goodfellow2014explaining} is a technique for training AI models in which adversarial inputs are generated for a model, and the model is then trained to give the correct outputs for those adversarial inputs.

Adversarial training can involve adversarial examples generated by human experts, human users, or other AI systems.

\rs{Robust overfitting in adversarial training} Adversarial training can be affected by robust overfitting, where the model’s robustness on test data decreases during further training, particularly after the learning rate decay. This issue has been consistently observed across various datasets and algorithms in adversarial training settings \cite{rice2020overfitting, yu2022understanding}. Robust overfitting can affect the model's ability to generalize effectively and reduce its resilience to adversarial attacks.

\rmmEX{Robustness certificates}
A model can be certified to withstand adversarial attacks given specific datapoint constraints, model constraints, and attack vectors \cite{raghunathan2018certified, lou2024cr}. Certification means that it can be both analytically proven and shown empirically that the model will withstand such attacks up to a certain threshold. 

Currently, robustness certification methods are limited to certifying against attacks via manipulation of pixels on specific $\ell^p$ norms, canonically the $\ell^2$ (Euclidean) norm, up to a certain neighborhood radius.

\rs{Robustness certificates can be exploited to attack the models} The knowledge of robustness certificates, including the area of the region for which model predictions are certified to be robust, can be used by an adversary to efficiently craft attacks that succeed just outside the certified regions \cite{cullentu}.

\rs{Poor model confidence calibration} Models can be affected by poor confidence calibration \cite{guo2017calibration}, where the predicted probabilities do not accurately reflect the true likelihood of ground truth correctness. This miscalibration makes it difficult to interpret the model’s predictions reliably, as high accuracy does not guarantee that the confidence levels are meaningful. This can cause overconfidence in incorrect predictions or underconfidence in correct ones.

\rmm{Calibrated confidence measures for model predictions} Incorporating calibrated confidence measures alongside a model’s predictions and standard performance metrics, such as accuracy, can help users identify instances of overconfidence in incorrect predictions or underconfidence in correct ones \cite{guo2017calibration}. These additional measures can provide users with more information to better interpret the model’s decisions and assess whether its predictions can be trusted.

\rmm{Incorporating the estimation of atypical input samples or classes for better model reliability} Incorporating the estimation of rare atypical input samples or classes might improve a model’s reliability, both with respect to its predictions and confidence calibration. Model predictions for rare inputs and classes may have a tendency of being overconfident and have worse accuracy scores \cite{yuksekgonul2024beyond}. For LLMs, the negative log-likelihood can be used as an atypicality measure. For discriminative models, Gaussian Mixture Models can be employed to estimate conditional and marginal distributions, which are then used in atypicality measurement.

\subsection{Fine-tuning-related}

\rs{Ease of reconfiguring GPAI models} GPAI models are often easily reconfigured for various use cases or have competencies beyond the intended use \cite{gade2023badllama, yang2023shadow}. They can be performed either by changing the weights of the model (e.g., fine-tuning) or by modifying only the model inputs (e.g., prompt engineering, jailbreaking, retrieval-augmented generation). Reconfiguration can be intentional (with the help of adversarial inputs) or unintentional (from unanticipated inputs to the model). 

\rs{Unexpected competence in fine-tuned versions of the upstream model} Downstream deployers may often fine-tune a GPAI model with specific deploy-\hspace{0pt}ment-related datasets, to better suit the task. Fine-tuned upstream models can gain new or unexpected capabilities that the underlying upstream models did not exhibit \cite{todd2023level, luo2024taiyi, muraoka2023cross}. These new capabilities may be unanticipated by the original model developer. 

\rs{Harmful fine-tuning of open-weights models} Models with publicly available weights can be fine-tuned for harmful activities by bad actors, using significantly fewer resources (in terms of time and money) compared to the original training cost \cite{lermen2023lora, gade2023badllama}.

\rs{Fine-tuning dataset poisoning} A deployer can poison the dataset used during the fine-tuning process \cite{jiang2023forcing} to induce specific, often malicious, behaviors in a model. This can be performed without having access to the model’s weights. This poisoning can be difficult to detect through direct inspection of the dataset, as the manipulations may be subtle and targeted.

\rmm{Data cleaning} Providers can filter out the training dataset via multiple layered techniques, ranging from rule-based filters to anomaly detection via data point influence or statistical anomalies of individual data points \cite{wang2022poisoning}.

For example, a data cleaning procedure can involve the use of filename checkers to detect duplicates or wrongly formatted data, which then moves to flagging the most influential data samples from the dataset via influence functions for anomaly detection.

\rmm{Internal data poisoning diagnosis} Providers can have an internal framework to identify what specific data poisoning attack their model may be a victim of based on a set of symptoms, such as analysis of target algorithm and architecture, perturbation scope and dimension, victim model, and data type \cite{chaalan2024path}. This framework includes known defenses against the diagnosed attack, which providers can then apply to the model.

\rs{Poisoning models during instruction tuning} AI models can be poisoned during instruction tuning when models are tuned using pairs of instructions and desired outputs. Poisoning in instruction tuning can be achieved with a lower number of compromised samples, as instruction tuning requires a relatively small number of samples for fine-tuning \cite{qiang2024learning, wan2023poisoning}. Anonymous crowdsourcing efforts may be employed in collecting instruction tuning datasets and can further contribute to poisoning attacks \cite{shu2023exploitability}. These attacks might be harder to detect than traditional data poisoning attacks.

\rs{Excessive or overly restrictive safety-tuning} Excessive safety training or safety tuning can impair the performance of AI systems, leading to overly cautious behavior. As a result, these systems may refuse to answer entirely safe prompts which are partially similar to harmful ones \cite{bianchi2023safety}.

\rs{Degrading safety training due to benign fine-tuning} When downstream providers of AI systems fine-tune AI models to be more suitable for their needs, the resulting AI model can be more likely to produce undesired or harmful outputs (as compared to the non-fine-tuned model), even if the fine-tuning was done with harmless and commonly used data \cite{qi2023fine}.

\rmmEX{Tamper-resistant safeguards for open-weight models} Training and implementing safeguards can improve the robustness of open-weight models against modifications from fine-tuning or other methods to change the learned weights of the models, especially those aimed at removing safety restrictions. These safeguards can be resilient even after extensive fine-tuning, ensuring that the model retains its protective measures \cite{tamirisa2024tamper}.

\rs{Catastrophic forgetting due to continual instruction fine-tuning} Catastrophic forgetting occurs when a model loses its ability to retain previously learned tasks (or factual information) after being trained on new ones. In language models, this can occur due to continual instruction tuning. 

This tendency may become more pronounced as the model’s size increases \cite{luo2023empirical}.

\section{Model Evaluations}
\label{Evaluations}

This section catalogs the risk sources and risk management measures related to model evaluations (often called evals). We categorize them into the following groups: general evaluations, benchmarking, red teaming, auditing, and interpretability/explainability. The subsection on general evaluations consists of items that are common to various evaluation techniques, while the other subsections are specific to their respective evaluation types.

\subsection{General Evaluations}

\rmm{Frequent testing when scaling model or dataset} Testing models after significant increases in compute, data, or model parameters. Even relatively small changes to model or dataset size can introduce new properties (“emergent abilities”) and failure modes. Identifying them early can prevent the models from being released prematurely \cite{anthropic2023RSP, barrett2023ai}.

\rmmEX{Using an AI model to evaluate AI model outputs} In cases where the outputs of AI models cannot be easily evaluated, AI models can be used to evaluate their outputs or the outputs of other AI models \cite{goodfellow2020generative, bai2022training, bai2022constitutional, irving2018ai}. The evaluations can then provide a training signal to improve the original model’s performance or offer explanations of the output for human users.

\rs{Incorrect outputs of GPAI evaluating other AI models} When an LLM is configured to evaluate the performance of another model or AI system, it may produce incorrect evaluation outputs \cite{liu2023evaluate, panickssery2024llm}. For example, it may give a higher rating to a more verbose answer or an answer from a particular political stance. If an LLM-based evaluation is integrated into the training of a new model, the trained model could develop in a way that specifically finds and exploits limitations in the evaluator’s metrics.

\rs{Limited coverage of capabilities evaluations} GPAI model developers might run capabilities evaluations to determine whether it has dangerous or dual-use capabilities, and then decide whether it is safe to deploy. Such capabilities evaluations can fail to demonstrate all the capabilities of a model. For example, evaluations may miss certain capabilities that are difficult to assess, prohibitively costly to verify, or obscured by the model’s tendency to refuse responses due to safety training, even if it possesses some of these capabilities.

\rs{Difficulty of identification and measurement of capabilities} The capabilities of general-purpose AI systems can be difficult to measure, compared to the capabilities of more limited and fixed-purpose AI systems. This is in part due to a broader distribution of potential risks, a lack of well-defined metrics to evaluate these risks, and risks from unpredictable (or emergent) AI model properties. 

Emergent properties are properties that cannot be predicted by extrapolating benchmark performance from smaller AI models, and may arise during training or deployment \cite{dong2023abilities, du2024understanding}. Such properties are more likely with large models. They can be especially difficult to detect since there are a large number of possible emergent properties of trained AI systems, but there does not currently exist a principled way of discovering the properties of a system from its components and weights alone.

For example, a more advanced general-purpose AI system can generate images of superior quality by a specialized prompt compared to generation without one, a property that did not exist in smaller or less advanced models of this type \cite{hendrycks2021unsolved}. There is currently no systematic method to identify such behaviors.

\rmmEX{GPAI models explaining model outputs in a zero-sum debate game} Debate is a technique that aims to produce reliable explanations of AI model outputs that are too complicated for humans to understand, by letting two GPAI models role-playing in a debate produce an explanation in a dialogue \cite{khan2024debating}.

For example, an AI model may produce an output which is time-consuming for humans to verify as doing so may require going through extensive sources.

Given such an output, the developer can use two natural language AI systems in an adversarial two-player setup to explain the output. These two natural language AI systems can be copies of the AI model that produced the output.

In this setup, one AI system gives a short explanation of the output. The second AI system responds to this explanation with a counter-explanation or an argument why the first explanation was not correct. This continues for a fixed number of turns, with the two AI systems pointing out inconsistencies in each others' explanations.

After this sequence of statements, a human or an AI judge evaluates the explanations of both AI systems to determine which one is more convincing. The results of this debate can then be used for further training of the models via reinforcement learning, where outputs that are more truthful and convincing arguments and explanations are positively reinforced, while misleading or false outputs are negatively reinforced.

\rs{ Self-preference bias in AI models} AI models may be prone to self-preference bias, where they favor their own generated content over that of others \cite{panickssery2024llm, laurito2024ai}. This bias becomes particularly relevant in self-evaluation tasks, where a model assesses the quality or persuasiveness \cite{durmus2024persuasion} of its own outputs, or in model-based evaluations more broadly. This bias can result in models unfairly discriminating against human-generated content in favor of their own outputs.

\rs{Inaccurate measurement of model encoded human values} There is a lack of robust frameworks for understanding and evaluating if the output of AI systems robustly conforms to human values, as opposed to if the systems have learned to produce outputs that are only partially correlated with them (i.e., mimicking) \cite{anwar2024foundational}. Additionally, outputs by AI models often do not perfectly reflect the representation of human values learned by the model, and it is not known how these values evolve and transition across different stages of model training and deployment.

Such evaluations may be especially challenging with LLMs that adopt different personas with different behaviorial patterns, where they do not consistently conform to certain human values.

\rs{Biased evaluations of encoded human values} Encoded human values in AI models that are easier to evaluate might be preferred for inclusion in evaluations over those that are more difficult to measure \cite{anwar2024foundational}. This might come at the expense of more desirable but harder-to-quantify values. This bias can lead to an imbalance, where easier-to-measure values dominate the evaluation process, while other important values are underrepresented.

\rs{AI outputs for which evaluation is too difficult for humans} When AI models are trained through evaluation with human feedback, such as reinforcement learning from human feedback, their outputs can be challenging to assess, as they may contain hard-to-detect errors or issues that only become apparent over time.

The human evaluator can rate incorrect outputs positively or similar to correct outputs. This can lead to the model learning to produce subtly incorrect or harmful outputs, such as code with software vulnerabilities, or politically biased information. In extreme cases where a model is deceiving users, complicated outputs can contain hidden errors or backdoors. 

For example, this can occur if an AI model is tasked with outputting a quarterly business plan whose quality will only be clear after the end of the quarter. Reaching consensus among experts when evaluating the business plan for efficacy may not happen even after long deliberation due to long-term uncertainty or unexpected events that affect plan efficacy.

\subsection{Benchmarking} 

\rmm{Benchmarking} Benchmarking is an evaluation method where different models are compared against a standardized dataset or a predetermined task. It allows comparison both across different models and over time, providing a reference point for model assessment. Benchmarks are usually open, where their question-answer pairs are publicly available \cite{zhou2023don}.

\rmm{Test robustness of GPAI system on relevant benchmarks} Various benchmarks \cite{croce2020robustbench, zhu2024promptrobust} have been developed to assess the robustness of GPAI systems when deployed in environments or scenarios that differ from their training conditions. These benchmarks typically evaluate the model’s ability to handle variations in inputs, unexpected data distributions, or adversarial examples, aiming to ensure reliable performance outside the original training domain.

\rmm{Frequent benchmarking to identify when red teaming is needed} Benchmarks, once created, are inexpensive to apply but may be less informative than red teaming. One reason is that sensitive data (e.g., relating to CBRN-related capabilities) cannot be included in the public questions and answers of benchmarks. On the other hand, red teaming can be more accurate given participants with diverse attack strategies, but it requires more resources to execute than benchmarking. If there is a correlation between benchmarking and red-teaming scores, then employing frequent benchmarking during the development of the model can identify arising vulnerabilities and inform the developers when more thorough red teaming is required \cite{barrett2024benchmark}.

Benchmarks can act as early warning signs of a larger issue, and red teaming can then be employed to investigate the severity and extent of such an issue.

\rs{Benchmark leakage or data contamination} Benchmark leakage \cite{zhou2023don, xu2024benchmarking, xu2024benchmark, ravaut2024much} can happen when an AI model is trained or fine-tuned with evaluation-related data. This can lead to an unreliable model evaluation, especially if the data contains question-answer pairs from benchmarks.

\rs{Raw data contamination} This type of contamination \cite{sainz2023nlp} occurs when the raw and unlabeled data of a benchmark is used as part of the training set. Such data may not be properly formatted and may contain noise, especially if the contamination happens before the data is pre-processed into the benchmark. If this contamination occurs, it could cast doubt on the few-shot and zero-shot performance of the model on that benchmark.

\rs{Cross-lingual data contamination} Models that have been trained on data encoded in multiple languages, such as LLMs trained on web-crawled data, may contain contamination that is obscured by translation \cite{yao2024data}. The most basic form of this is when a benchmark is translated to another language and then fed to the model as training data. The fact that the benchmark is translated before becoming training data can obscure the contamination from detection methods, giving false assurance that the model has generalized on the capabilities that the benchmark tests for.

\rs{Guideline contamination} Guideline contamination refers to scenarios where instructions for the collection, annotation, or use of the dataset are exposed to the model \cite{sainz2023nlp}. These instructions may contain explicit data-label pairs that can improve the model’s capabilities for the task.

For example, for text-based models, this can include prompts used to generate synthetic data, as well as instructions for evaluators on the coverage and method of their evaluations of the model.

\rs{Annotation contamination} Annotation contamination refers to scenarios where the model is exposed to the benchmark labels during training \cite{sainz2023nlp}. This type of contamination can make the model learn the acceptable distribution of outputs. Combining this with raw data contamination of the test split, any evaluation made with the benchmark is invalidated because the entire test split is essentially leaked to the model. 

\rs{Post-deployment contamination} Once a model is deployed, it can be exposed to benchmark data provided by the users \cite{jacovi2023stop, sainz2023nlp}. The model may then be further trained by these user inputs containing benchmark data.

\rmm{Contamination detection} Contamination detection refers to techniques for assessing whether and to what extent a given model has benchmark data in its training dataset \cite{ravaut2024much, sainz2023nlp}. This can involve a set of technical and regulatory interventions to prevent or identify a model trained on contaminated data.

For example, with web-crawled data, contamination detection can involve comparing the data’s web sources against a dynamic, publicly available blocklist of websites known to generate new benchmarks. Additional measures may include excluding data with improper metadata from the training dataset and conducting overlap analyses between the training data and all known standard benchmark datasets.

\rmm{Preventing or mitigating data contamination and leakage} Developers of GPAIS and creators of benchmarks can take actions to prevent AI models from being trained on contaminated or leaked data, or mitigate such data contamination and leakage. 

For example, developers of AI models can try to find and remove contaminated or leaked data from the training corpus, and creators of benchmarks can help them by adding globally unique “canary strings" to the documents containing their benchmarks, which makes them easier to find \cite{srivastava2022beyond}. More involved interventions by benchmark-creators include restricting access to benchmarks over an API, or continually updating benchmarks to focus on recent data.

\rmmEX{Tracking benchmark leakage} Constant monitoring and documentation of benchmark leakage can help with the early detection of benchmark leaks \cite{xu2024benchmarking, jacovi2023stop}.

\rmm{Reporting data decontamination efforts} Decontamination analysis may involve comparing the training dataset with the benchmark data, and publishing a report with statistics such as data overlap \cite{zhou2023don}. Especially for already trained models, including contamination statistics can allow experts to reweigh the success of the model on affected benchmarks.

\rmm{Avoiding benchmark data with publicly available solutions and releasing contextual information for internet-derived data} Evaluators can improve the integrity of benchmarks by avoiding data with publicly available solutions. When using internet-derived data, supplementing it with contextual information (such as entity linking) may help to better document and assess the integrity of the collected data \cite{jacovi2023stop}. This could help in detecting instances where contextual information (which may have been included in the training data) might inadvertently reveal details about the solution, ensuring that the data is suitable for accurate benchmarking.

\subsubsection{Benchmark Inaccuracy}

\rs{Benchmarks may not accurately evaluate capabilities} Benchmarks of AI systems can both underestimate and overestimate the capabilities of those AI systems.

Underestimates can happen if an evaluation is not comprehensive enough, if the benchmark is saturated by existing models, or if the capabilities in question depend on a complicated setup, such as realistic computer programming tasks.

Overestimates of capabilities can occur if an AI system is trained or fine-tuned on the contents of the benchmark, leading to overfitting.

\rs{Benchmark saturation} Benchmark saturation refers to benchmarks reaching their evaluation ceiling. The tendency towards benchmark saturation has been demonstrated in various benchmarks \cite{barbosa2022mapping}. When benchmarks reach or are close to saturation, they stop being effective measures for new models, as more nuanced capability gains might not be detected.

\rmm{Statistical data quality reports for benchmarks} If a benchmark dataset is too large to allow for the identification and removal of all flawed instances, statistical reports on the data composition can be added. Random sampling of benchmark data points can be performed to evaluate and report the frequency and types of errors found \cite{davis2023benchmarks}.

\subsubsection{Benchmark Limitations}

\rs{Insufficient benchmarks for AI safety evaluation} Benchmarks dedicated to measuring the performance of AI systems (e.g., on programming or math tasks) are more well-developed than those for assessing safety and harms in AI systems \cite{zhang2023safetybench}. This gap can lead to AI systems excelling in specific tasks while exhibiting harmful behaviors that go undetected. More safety-related evaluation datasets can help in identifying previously overlooked undesirable model behaviors. 

Additionally, statistical analysis of safety-related benchmarks shows that high scores correlate substantially with model performance \cite{ren2024safetywashing}. This may allow for performance improvements to be misrepresented by model providers as safety improvements.

\rs{Underestimating capabilities that are not covered by benchmarks} A lack of test coverage by benchmarks on specific abilities of a model can obscure the model’s capabilities from both the developer and the user \cite{rameshcompositional}. This can lead to a false sense of safety and trust due to a lack of understanding of the model's limitations.

\subsubsection{Good Benchmarking Practices}

\rmm{Informative and powerful benchmarks} Developers of GPAI systems can select benchmarks that are difficult enough to be informative about the capabilities of their AI systems, and cover a large spectrum of domains in order to signal areas where the GPAI system is performing poorly \cite{liang2022holistic}.

Suitable benchmarks contain no label errors, are not vulnerable to being benchmark contaminants, and are often audited by independent domain experts if they contain domain-specific questions. For multimodal GPAI systems, good benchmarks cover every modality, especially the interaction of different modalities.

\rmm{Benchmark dataset auditing}
Auditing benchmark datasets allows for verification of the utility and limitations of the datasets \cite{rajore2024truce}. This allows the provider to more accurately measure AI model capabilities and safety. Auditing includes the evaluation of such datasets by independent third-party organizations and the release of benchmark dataset metadata to the auditors.

\rmm{Dynamic benchmarking} Dynamic benchmarks are benchmarks that can be continuously updated with new human-generated data. By having one or more target models “in the loop,” examples for benchmarking can be generated with the intent of fooling these target models, or to assess if these models express an appropriate level of uncertainty \cite{kiela2021dynabench}. As the dataset in the benchmark grows, previously benchmarked models can also be reassessed against the updated dataset to reflect its performance in a more representative manner. 

Examples of such dynamic benchmarks include DynaSent \cite{potts2020dynasent} for sentiment analysis, LFTW \cite{vidgen2020learning} for hate speech, and Human-Adversarial VQA \cite{sheng2021human} for images.

\subsection{Red Teaming}

\rmm{Red teaming for GPAI system evaluation} Red teaming refers to simulated adversarial attacks performed to identify and evaluate the model’s vulnerabilities as well as its in-domain and out-of-domain performance.

\rmm{Red team access to the final version of a model pre-deployment} Granting red teams access to the final pre-release version of the model can help with identifying potentially dangerous model properties. These properties might not be identified if red teaming is only performed on earlier versions of the model, as late fine-tuning procedures may introduce new vulnerabilities.

Red-teaming AI models before they are released to the public can reduce the model non-decomissionability risk. The model’s release can be postponed or even prevented if previously unidentified flaws are detected during the testing \cite{dubey2024llama}.

\rmm{Scope red-teaming activities based on deployment context} Red-teaming activities can be tailored and compared based on the specific deployment circumstances of an AI system. This involves adapting the scope, depth, and focus of red-teaming efforts to match the intended use case, potential risks, and operational context of the AI system.

Points of consideration include: 

\begin{itemize}
  \item The diversity of potential users and use cases
  \item The sensitivity and impact of the application domain
  \item The scale of deployment and potential reach
  \item Known vulnerabilities or concerns specific to the model or similar systems
\end{itemize}

\rmmEX{Red teaming to test the resilience of open-weights models to fine-tuning} Before the release of open-weights models, red teamers can test the resilience of safety training against fine-tuning. Safety training may be partially or fully overridden by fine-tuning intentionally (e.g., by malicious actors) or unintentionally (e.g., by fine tuning an AI model for a specific use case) \cite{rosati2024immunization}.

\subsection{Auditing}

\rs{Conflicts of interest in auditor selection} Conflicts of interest can arise if there is no independence in the auditor selection process or if the auditors are closely associated with the developer \cite{longpre2024safe, raji2022outsider}. In such cases, the conflict of interest can appear even if third-party evaluators are involved. In the case of external auditing, the potential candidates might be selected from a narrow group of auditors, or have conflicting financial incentives for whether to report model shortcomings publicly.

\rs{Auditor capacity mismatch} Auditors may not be able to address all of the specific safety, performance, or validation needs. Reports of passing audits may be more inclusive than can be justified due to a lack of knowledge of specific risks and how they can be tested, or a lack of capacity to perform sufficiently rigorous testing. 

\rs{Auditor failure} Auditors may not publicly disclose risks they find, may be required to not publicize shortcomings, or may not receive sufficient cooperation from the relevant internal parties.

\rmm{Pre-deployment access by third-party auditors} Prior to full deployment of general-purpose AI models, a group of third-party auditors who are not selected by the GPAI model provider could get early access to AI models in order to evaluate them from a variety of different perspectives and with diverse interests \cite{bommasani2021opportunities, raji2022outsider}.

This prevents cases where the developers of AI models select auditors that are especially favorable to the developers, which could result in biased or incomplete evaluations, or contribute to an unjustified public perception of the capabilities and risks of the model.

\rmm{Audits with specific scoped goals} Audits of AI systems will be easier to perform, and have clearer results if the scope and goals of the evaluation are formulated as precisely as possible \cite{raji2022outsider}, or have a connection to concrete existing policy. 

\subsection{Interpretability / Explainability}

\rs{Misuse of interpretability techniques}  Interpretability techniques, by enabling a better understanding of the model, could potentially be used for harmful purposes. For example, mechanistic interpretability could be used to identify neurons responsible for specific functions, and certain neurons that encode safety-related features may be modified to decrease its activation or certain information may be censored \cite{bereska2024mechanistic}.

Furthermore, interpretability techniques can be used to simulate a white-box attack scenario. In this case, knowing the internal workings of a model aids in the development of adversarial attacks \cite{bereska2024mechanistic}.

\rs{Misunderstanding or overestimating the results and scope of interpretability techniques} The results of explainability techniques are not free of bias and require careful interpretation. Users might develop a false sense of security or reliability if the resulting explanations align with their initial beliefs, leading to confirmation bias and an overestimation of abilities of these techniques \cite{bereska2024mechanistic}. 

For example, it has been demonstrated that some interpretability techniques in computer vision display object edges as salient in a heatmap, regardless of the underlying model \cite{adebayo2018sanity}. This might create a false sense of confidence in the interpretability technique. 

\rs{Adversarial attacks targeting explainable AI techniques} Adversarial attacks can affect not only the model’s output but also its corresponding explanation. Current adversarial optimization techniques can introduce imperceptible noise to the input image, so that the model’s output does not change but the corresponding explanation is arbitrarily manipulated \cite{dombrowski2019explanations}. Such manipulations are harder to notice, as they are less commonly known compared to standard adversarial attacks targeting the model’s output. 

\rs{Biases are not accurately reflected in explanations} Existing explainability techniques can be insufficient for detecting discriminatory biases. Manipulation methods can hide underlying biases from these techniques, generating misleading explanations \cite{slack2020fooling, lakkaraju2020fool}. Such explanations exclude sensitive or prohibitive attributes, such as race or gender, and instead include desired attributes, even though they do not accurately represent the underlying model.

\rmm{Evaluating explainability method sensitivity to data inputs and model parameters} Model parameter and data randomization tests can be employed as sanity checks \cite{adebayo2018sanity} to determine the relationship between a model, its inputs, and the explainability method. If an explanation is independent from or insensitive to the underlying model or the input data, it would indicate a lack of reliability of the explainability method.

\rs{Model outputs inconsistent with chain-of-thought reasoning} Chain-of-thought reasoning is sometimes employed to get a better understanding of the model’s output, where it encourages transparent reasoning in text form. However, in some cases, this reasoning is not consistent with the final answer given by the AI model, and as such does not give sufficient transparency \cite{lanham2023measuring}.

\rmmEX{Enforcing model output interpretability post-training} While a resulting trained model may be opaque with respect to its predictions, the final output in a system that involves the model can nonetheless be completely interpretable.

For example, a neuro-symbolic system for robot navigation can use a language model to generate potential navigation plans, then have a deterministic solver simulate executing valid plans \cite{lyu2023faithful}. The optimal plan found is executed in the real world and is interpretable.

\rs{Encoded reasoning} Models can employ steganography techniques to encode their intermediate reasoning steps in ways that are not interpretable by humans \cite{roger2023preventing}. Since encoded reasoning can improve model performance, this tendency might naturally emerge and become more pronounced with more capable models.

\rmm{Paraphrasing to reduce hidden information} Paraphrasing can mitigate steganography and encoded reasoning by reducing the physical size of the hidden information encoded in text \cite{roger2023preventing}.

\rmm{Testing for erroneous or irrelevant features through concept learning} Interpretability techniques, particularly concept learning \cite{fong2018net2vec}, can be used to test whether a model is learning erroneous features or relying on irrelevant features in its predictions. This can help identify and mitigate potential risks associated with incorrect or non-informative features influencing the model's outputs.

\rmm{Dashboard of model properties} A dashboard \cite{chen2024designing} displays all the relevant information about the model’s internal state and the model’s physical properties to the user. It is used to ensure that the user is informed about factors that influence the behavior of the model and to ensure that the user maintains control over the model. Allowing only the user to access the dashboard can aid in information asymmetry between the user and model, thus supporting the user oversight over the model.

Examples of the model's internal state include its representations of the world, representation of users, and the strategies it is currently pursuing; while examples of the model's physical properties include the model’s compute and energy consumption, physical storage occupied, and physical networks it is connected to.

\rmmEX{Modification of model internal representation} Model providers can modify the internal representation of the model \cite{meng2022mass, zou2023representation} up to the granular level and in consultation with other tools that aid in understanding what the internal representation of the model is. 

For example, if a model that is meant to give factual health data learned an incorrect fact, the model provider can use Linear Artificial Tomography (LAT) \cite{zou2023representation} to identify the representations responsible for the incorrect fact, and then modify that representation via modifying individual weights, or modify the entire representation itself by modifying entire layers.

\section{Attacks on GPAIs / GPAI Failure Modes}
\label{Attacks}

This section catalogs the risk sources related to GPAI failure modes or attacks targeting GPAIs. Many of these apply mainly to LLM-based GPAIs, which share some common failure modes such as jailbreaks and trojans. These vulnerabilities often extend beyond GPAIs and fall into the broader field of adversarial machine learning. However, additional vulnerabilities may arise with the introduction of new modalities, longer context windows, or different encodings.

\rs{Jailbreak of a model to subvert intended behavior} A jailbreak is a type of adversarial input to the model (during deployment) resulting in model behavior deviating from intended use. Jailbreaks may be generated automatically in a “white box” setting, where access to internal training parameters is required for creation and optimization of the attack \cite{zou2023universal}. Other attacks may be “black box” - without access to model internals. In text based generative models, jailbreaks may sometimes be human-readable, with the use of reasoning or role-play to “convince” the model to bypass its safety mechanisms \cite{yu2024don}.

\rs{Jailbreak of a multimodal model} Current generation multimodal (e.g., vision and language) GPAI models are vulnerable to adversarial jailbreak attacks. These attacks can be used to automatically induce a model to produce an arbitrary or specific output with high success rate \cite{yin2024vlattack}. Multimodal jailbreaks can also be used to exfiltrate a model’s context window or other model internals \cite{bailey2023image}.

\rs{Transferable adversarial attacks from open to closed-source models} In some cases, an adversarial attack developed for an open-weights and open-source model (where the weights and architecture are known - a “white box” attack) can be transferable to closed-source models, despite the defenses put in place by the closed-source model provider (such as structured access). These adversarial attacks can be generated automatically \cite{zou2023universal}.

\rs{Backdoors or trojan attacks in GPAI models} Backdoors can be inserted into GPAI models during their training or fine-tuning, to be exploited during deployment \cite{shi2023badgpt, li2024unveiling}. Attackers inserting the backdoor can be the GPAI model provider themselves or another actor (e.g., by manipulating the training data or the software infrastructure used by the model provider) \cite{xu2023instructions}. Some backdoors can be exploited with minimal overhead, allowing attackers to control the model outputs in a targeted way with a high success rate \cite{hubinger2024sleeper}.

\rs{Text encoding-based attacks} Various new or existing text encodings, such as Base64, can be employed to craft jailbreak attacks that bypass safety training \cite{anwar2024foundational}. Low-resource language inputs also appear more likely to circumvent a model’s safeguards \cite{yong2023low}. Since safety fine-tuning might not involve this encoding data or may only do so to a limited extent, harmful natural language prompts could be translated into less frequently used encodings \cite{wei2024jailbroken}.

\rs{Vulnerabilities arising from additional modalities in multimodal models} Additional modalities can introduce new attack vectors in multimodal models as well as expand the scope of the previous attacks, ranging from jailbreaking to poisoning \cite{anwar2024foundational}. Typically, different modalities have different robustness levels, allowing malicious actors to choose the most vulnerable part of the model to attack \cite{li2024images, shayegani2023survey}. 

\rs{Vulnerabilities to jailbreaks exploiting long context windows (many-shot jailbreaking)} Language models with long context windows are vulnerable to new types of exploitations that are ineffective on models with shorter context windows. While few-shot jailbreaking, which involves providing few examples of the desired harmful output, might not trigger a harmful response, many-shot jailbreaking, which involves a higher number of such examples, increases the likelihood of eliciting an undesirable output. These vulnerabilities become more significant as context windows expand with newer model releases \cite{anil2024many}.

\rs{Models distracted by irrelevant context} Models can easily become distracted by irrelevant provided information (such as “context” in LLMs), leading to a significant decrease in their performance after introducing irrelevant information. This can happen with different prompting techniques, including chain-of-thought prompting \cite{shi2023large}.

\rs{Knowledge conflicts in retrieval-augmented LLMs} AI models can be particularly sensitive to coherent external evidence, even when they come into conflict with the models’ prior knowledge. This may lead to models producing false outputs given false information during the retrieval-augmentation process, despite only a relatively small amount of false information input that is inconsistent with the model’s prior knowledge trained on much larger amounts of data \cite{xie2023adaptive}.

\rs{Lack of understanding of in-context learning in language models
} In-context learning allows the model to learn a new task or improve its performance by providing examples in the prompt, without changing its weights \cite{kaplan2020scaling}. Even though this technique is highly effective, its working mechanism is not well understood. Since many potential misuses are directly related to prompting, it becomes difficult to guarantee safety when the exact mechanism of in-context learning is not fully investigated \cite{anwar2024foundational}.

For example, in-context learning has been used to re-learn forbidden tasks in models that have been fine-tuned not to engage in the forbidden behavior \cite{xhonneux2024context, anil2024many}.

\rs{Model sensitivity to prompt formatting} LLMs can be highly sensitive to variations in prompt formatting, such as changes in separators, casing, or spacing. Even minor modifications can lead to significant shifts in model performance, potentially affecting the reliability of model evaluations and comparisons. This sensitivity persists across different model sizes and few-shot examples \cite{sclar2023quantifying}.

\rs{Misuse of AI model by user-performed persuasion} AI models can be influenced to accept misinformation through persuasive conversations, even when their initial responses are factually correct. Multi-turn persuasion can be more effective than single-turn persuasion attempts in altering the model’s stance \cite{xu2023earth}.

For example, a Question \& Answering AI model trained to be helpful and agreeable to users can be persuaded to produce obviously incorrect answers when explicitly prompted by the user. This can aid in misuse, such as persuading the model to perform a harmful phishing task \cite{wang2023can}.

\section{Agency}
\label{Agency}

This section catalogs the risk sources and risk management measures related to agentic AI systems. We categorize these into the following groups: goal-directedness, deception, situational awareness, self-proliferation, and persuasion. These risk items are related to behaviors associated with agentic systems using a GPAI as a base model or other component that is tasked with achieving an objective by manipulate its environment. They should not be confused with autonomy, where an AI system is free to take actions without full human supervision. Risks related to autonomous AI systems are further discussed in Appendix \ref{sec:autonomous}.

\subsection{Goal-Directedness}
\rs{Specification gaming} AI systems can achieve user-specified tasks in undesirable ways unless they are specified carefully and in enough detail. AI systems might find an easier unintended way to accomplish the objective provided by the user or developer, so that the actions by the AI system taken during its execution are very different from what the user expected \cite{feldt1998generating, skalse2022defining}. This behavior arises not from a problem with the learning algorithm, but rather from the misspecification or underspecification of the intended task, and is generally referred to as specification gaming \cite{chu2017cyclegan}.

\rs{Reward or measurement tampering} Measurement and reward tampering occur when an AI system, particularly one that learns from feedback for performing actions in an environment (e.g., reinforcement learning), intervenes on the mechanisms that determine its training reward or loss. This can lead to the system learning behaviors that are contrary to the intended goals set by the developer, by receiving erroneous positive feedback for such actions. 
This has two main forms:

\begin{enumerate}
    \item Measurement tampering: The AI system interferes with the sensors or data collection processes that measure its performance, causing inaccurate feedback \cite{roger2023benchmarks}. This is especially applicable in embodied AI systems that affect the physical world.
    \item Reward tampering: The AI system directly modifies its reward function or the process that calculates rewards \cite{everitt2021reward}. This is especially applicable in non-embodied systems (e.g., coding assistants).
\end{enumerate}

Measurement tampering can be viewed as a subset of specification gaming, and it might affect more capable AI systems. 

\rs{Specification gaming generalizing to reward tampering} In some instances, specification gaming in a GPAI model can lead to reward tampering, without further training. This can mean that relatively benign cases of specification gaming (such as sycophancy in LLMs) can, if left unchecked, enable the model to generalize to more sophisticated behavior such as reward tampering \cite{denison2024sycophancy}.

\rs{Goal misgeneralization} Goal or objective misgeneralization is a type of robustness failure where an AI system appears to be pursuing the intended objective in training, but does not generalize to pursuing this objective in out-of-distribution settings in deployment while maintaining good deployment performance in some tasks \cite{shah2022goal, di2022goal}.

This behavior might not be detected in the training or even testing environments but can have negative outcomes during the deployment phase. In contrast to capability misgeneralization, where an AI system performs generally poorly under distributional shift, in goal misgeneralization scenarios the system might still efficiently perform different actions or tasks, just towards a wrong objective.

For example, a meeting scheduling chatbot may learn that the user prefers meetings at a restaurant, as opposed to online meetings: 

\begin{itemize}
    \item Before COVID-19, the user only scheduled meetings at restaurants.
    \item During COVID-19, the user might request online-only meetings.
\end{itemize}
At that point, the chatbot could misgeneralize and, instead of agreeing to schedule an online meeting, attempt to persuade the user that the restaurant is safe. In that case, the goal the chatbot learned is “schedule meetings at restaurants” instead of “schedule meetings at the user's preferred location” \cite{shah2022goal}.

While failing to perform the intended function, the chatbot is not exhibiting a standard robustness failure, as it shows competence in scheduling, persuasion, and meeting scheduling, but is directed towards an unintended goal.

\rmmEX{Incorporating goal uncertainty into AI systems to mitigate risky behaviors} Incorporating uncertainty into AI system goals can prevent rushed decision-making and incentivize such systems to gather additional information or to refer to human oversight when faced with ambiguity \cite{yohsua2024international}.

\subsection{Deception}
\rs{Deceptive behavior}Deceptive behavior of an AI system consists of actions or outputs of the AI that reliably mislead other parties, including humans and other AI systems. This behavior can result in the targeted parties becoming convinced of, and acting on, false information \cite{ngo2022alignment}.

Deceptive behavior can occur due to several different reasons, including \cite{park2024ai}:
\begin{enumerate}
    \item The developer trained, programmed, or configured the AI system to behave deceptively.
    \item In AI systems capable of planning, deceptive outputs arise when the behavior is optimal for the goals the AI systems have been configured or trained to achieve.
    \item The training data of the AI system contains repeated incorrect information, or the feedback from human raters on AI outputs is biased.
\end{enumerate}
An AI system may produce deceptive outputs because their learned world model is not an accurate model of the real world \cite{vincent2023microsoft}.

\rs{Deceptive behavior for game-theoretical reasons} An AI system can display deceptive behavior, such as cheating or bluffing, when engaging in such behavior is a good or optimal game-theoretical strategy to achieve the goals it has been configured to achieve.  This tendency can exist in AI systems designed to maximize reward or utility, whether these designs use machine learning or not. The use of deceptive strategies has been demonstrated in both narrow and general AI systems, in both game-playing systems and in systems not explicitly designed to treat humans as opponents, and in systems using both very simple machine learning (e.g., Q-learners) and very complex machine learning \cite{brown2019superhuman, meta2022human}.

\rs{Deceptive behavior because of an incorrect world model} AI systems can create deceptive outputs because their learned world model is not an accurate model of the real world \cite{vincent2023microsoft}.

\rs{Deceptive behavior leading to unauthorized actions} AI systems can create false or misleading claims that can lead to unauthorized actions, even in some cases violating the terms and conditions set by the model provider \cite{garcia2024air, achiam2023gpt}.

For example, an AI system can claim that it is not collecting data from its current interaction with the user, in line with the provider’s policies, but the system still stores the user's input without deleting it after the session. This harms both the user and the provider, as the provider is exposed to increased legal liability due to the model's actions.

\rmm{Evaluations for truthful outputs} AI systems can be evaluated for truthfulness when  answering questions, including in contexts where humans tend to give incorrect but widely-accepted answers (i.e., popular misconceptions). Evaluations detect incorrect facts learned about the world, inadequate capabilities of the AI system, and misleading outputs by the AI system \cite{lin2021truthfulqa}.

\rmmEX{Interpretability techniques that target deception} Interpretability techniques can be used for finding the root causes of outputs of an AI model that reliably lead to false beliefs for its users (e.g., deceptive behavior) \cite{shah2022goal}.

It is often difficult to distinguish a deceptive AI model from an honest AI model, since absence of deception and very sophisticated (hard to detect) deception may appear behaviorally similar. Interpretability techniques and tools can be used to detect whether AI model outputs arise from internal computations representing deception. This can apply in cases of purposely trained deception by the developer or if it emerges unintentionally during training.

These interpretability tools can come from mechanistic interpretability, such as identification of features involved in generating the outputs, or attribution of parts of the input most important in generating the output \cite{macdiarmid2024simple}.

\subsection{Situational Awareness}
\rs{Situational awareness in AI systems} Situational awareness in GPAI systems refers to the ability to understand its context, environment, and use this to inform action. This can range from basic environmental mapping and trajectory estimation (as in a robot vacuum cleaner) to sophisticated understanding of its training, evaluation, or deployment status. In more advanced systems this may enable undesired behavior, such as deceptive behavior during evaluations, or persuasion during deployment.

For a GPAI model, the types of awareness can include \cite{ngo2022alignment, berglund2023taken}:
\begin{itemize}
    \item Environment: Understanding and modeling the physical or digital environment in which it operates.
    \item Context: Identifying whether it is in training, testing, evaluation, or deployment, as well as knowledge of its capabilities, limitations, and techniques used in training. 
    \item User: Understanding user expectations, inferring personal characteristics (e.g., age, political leaning, education), and expected responses to the AI's actions. 
\end{itemize}

\rs{Strategic underperformance on model evaluations} GPAI developers often run evaluations ofual-use capabilities to decide whether it is safe to deploy. In some cases, these evaluations may fail to elicit these capabilities, either due to benign reasons or strategic action - by either the developers, malicious actors, or arise unintentionally in the model during training \cite{greenblatt2023ai, jarviniemi2024uncovering}.

A GPAI model may strategically underperform or limit its performance during capability evaluations in order to be classified as safe for deployment. This underperformance could prevent the model from being identified as potentially dual use.

Some examples include:
\begin{itemize}
    \item During training involving user feedback (e.g., reinforcement learning from human feedback), an AI model may provide different answers to evaluators who introduce themselves as less educated, and therefore less able to judge accurately \cite{perez2022discovering}.
    \item Of particular concern is an AI system employing deception to manipulate performance evaluations, as has already occurred with some non-AI systems, such as in the Volkswagen emissions scandal \cite{crete2016volkswagen}.
\end{itemize}

\rmm{Benchmarks for Situational Awareness} A model is situationally aware if it internally represents that it is a machine learning model and if it can accurately infer or act on model-relevant facts - e.g., if it is currently in training, testing, evaluation or deployment, or the desired outcome of an evaluation.

Some benchmarks exist for situational awareness of AI models, which test whether the AI models can classify stereotypical inputs from training, testing, evaluation and deployment as such, and whether the AI model can use this information correctly to take actions in the world \cite{berglund2023taken, laine2023towards}.

\subsection{Self-Proliferation}

\rs{Self-proliferation} An AI system can self-proliferate if it can copy itself and its constituent components (including its model weights, scaffolding structure, etc.) outside of its local environment \cite{cohen2024here}. This can include the AI system copying itself within the same data center, local network, or across external networks \cite{kinniment2023evaluating}.

The self-proliferation of an AI system can include acquisition of financial resources to pay for computational resources via work or theft, the discovery or exploitation of security vulnerabilities  in software running on publicly accessible servers, and persuasion of humans \cite{anthropic2024claude, lu2024ai}.

Self-proliferation may be initiated by a malicious actor (e.g., by model poisoning), or by the model itself.

\rmmEX{Evaluating AI systems’ performance on self-proliferation-related tasks} To prevent AI systems from self-proliferating, developers of AI systems can evaluate those systems for their capability to engage in self-proliferation-related tasks. 

This type of evaluation might assess an AI system’s ability to replicate its components (including model weights, structural scaffolding, etc.) onto other local or cloud infrastructures prior to deployment. Additionally, it may test the system’s capacity to purchase cloud credits and configure virtual machines on a cloud platform. The evaluation could offer a predefined environment (such as a virtual container) to facilitate self-replication, providing access to the system’s own components, a network connection to a resource-equipped external computer, and other necessary resources \cite{anthropic2024claude, kinniment2023evaluating}.

Self-proliferation evaluations can be conducted in a secure environment to prevent a self-replicating AI system from affecting other computers.

\subsection{Persuasion}
\rs{Persuasive Capabilities} GPAI systems can produce outputs (such as natural language text, audio, or video) that convince their users of incorrect information. This can happen through personalized persuasion in dialogue, or the mass-production of misleading information that is then disseminated over the internet. The persuasive capabilities of GPAI models can sometimes scale with model size or capability \cite{breum2024persuasive, salvi2024conversational}. Persuasive models could have larger societal implications by being misused to generate convincing but manipulative or untruthful content.

\section{Deployment}
\label{Deployment}

This section catalogs the risk sources and risk management measures related to the model deployment stage. We categorize these into the following groups: risk assessment, post-deployment practices, and monitoring. The risk items in this section are most relevant to the deployer of an AI system, who may be deploying a GPAI model developed by a different entity. They include risk management practices to inform deployment decisions, processes related to the release of the AI system, actions relevant to systems that have already been deployed, and long-term monitoring measures for systems in operation. 

\subsection{Risk Assessment}

The risk assessment techniques below are generally accepted in other industries, and may be useful in GPAI systems. Some of these techniques are already being explored or used to a limited extent in the context of systemic risk management. 

\rmm{Demonstrating a “margin of safety” for the worst plausible system failures} Model developers can demonstrate that there is an acceptable “margin of safety” between the current version of the model and a plausible version with dangerous capabilities or potential system failures, whether these arise from the model itself or through scaffolding. This “margin of safety” can be tracked and evaluated based on the model’s performance on either component tasks or proxy tasks with varying levels of difficulty \cite{clymer2024safety}, and it is particularly relevant for general-purpose models with emergent properties, where some of the risks, use cases, and model capabilities may be unknown even at the time of deployment.

Margin of safety (also called “safety factor”) is a common practice in many industries - particularly in physical structures. It is common for this margin to be very conservative when feasible (e.g., 4+ in fasteners on critical structures). In situations where a high margin of safety is impractical, it may be supplemented by more frequent inspections and additional process controls. A lower safety factor can also be managed by adopting conservative assumptions regarding worst-case conditions.

\rmm{Employing qualitative assessments in difficult-to-measure domains} Qualitative evaluation can be used in cases when quantitative measurement is not feasible. This can give additional insights about the system which would not be available if no measurement was performed due to its difficulty \cite{vianello2023improving}.

\rmm{Scenario analysis} Scenario analysis involves development of several plausible future scenarios, where these scenarios may be generated from varying the assumptions of a small set of driving forces. The scenarios developed can be used to take further actions to improve overall preparedness \cite{koessler2023risk}. 

For example, scenarios to explore frontier AI risks can be developed, where they can involve different assumptions on factors such as AI capability, ownership, safety, usage, and geopolitical contexts, as well as its implications on key policy issues \cite{UK_risks}.

\rmm{Fishbone diagram} The fishbone diagram, or a “cause-and-effect diagram" \cite{koessler2023risk}, can be used to show the potential causes of an undesirable event. 

The diagram is created by first placing a specific risk event at the “head" of the diagram, typically facing the right. Then, to the left of the risk event, the “ribs" branch off from the “backbone" to represent major causes, which further branch into sub-branches to represent root-causes, extending to as many levels as required. This is typically done via backward-reasoning, where various potential causes are explored after the risk event has been selected for analysis using this method. 

For example, the risk event “AI systems generate toxic content" can be placed at the “head" of the diagram, where the branches may include causal factors like “AI trained on data containing toxicity" and “successful jailbreaking of AI despite fine-tuning."

\rmm{Causal mapping} Causal mapping is a technique used to explore and map complex interactions between cause and effect of risks. It involves coming up with potential events related to an undesirable issue, with each event represented by a text box, then clustering similar events according to themes, and finally drawing arrows to illustrate the causal relationship between the different events. The completed causal map can then be analyzed to identify central events, clusters of events, feedback loops, and other relevant patterns \cite{koessler2023risk}. 

For example, causal mapping can be used to explore factors that lead to high-level model capabilities (e.g., “machine intelligence”), and the nodes may include factors such as “concept formation" and “flexible memory," where certain nodes may be found to be especially crucial if they have more outgoing arrows connecting them to other nodes. 

\rmm{Delphi technique} The Delphi technique is a multi-round forecasting process based on a structured framework on collecting and collating multiple expert judgments. It brings the benefits of anonymous and remote participation which may result in increased likelihood estimation accuracy compared to merely averaging individual estimations or simple group discussions \cite{koessler2023risk}. 

Given a panel of experts, at each round, the experts are presented with an aggregated summary of the results from the previous round, and are then allowed to update their answers accordingly. The process ends when either a consensus is reached or the responses in later rounds no longer change significantly. This method enables elicitation of expert judgment while utilizing wisdom of the crowd in the process. 

A potential application of the Delphi technique is to solicit expert judgment on the likelihood of systemic risks from AI development, where crucial variables identified during each round of questionnaire can be further studied for the purpose of risk mitigation.

\rmm{Cross-impact analysis} Cross impact analysis is a forecasting methodology that analyzes the likelihood of a particular issue using expert analysis (i.e., Delphi technique) in combination with analysis of events correlated with the said issue. It involves decomposing an issue into discrete and correlated events, and then collecting expert opinion on each of those events. Analysis of each event from multiple viewpoints can yield potential future scenarios \cite{koessler2023risk}. 

For example, an issue may be “advances in AI," which can be broken down into two correlated events like “advances in hardware" or “advances in algorithms," where the likelihood of occurrences of each event can be estimated via the Delphi technique while taking into account their interactions with other events.

\rmm{Bow-tie analysis} Bow-tie analysis is a method to assess the utility of implemented controls against a particular risk event. It involves centering the unwanted risk event within a diagram. On the left, the factors that can cause the event are listed, followed by the controls that will prevent or minimize the likelihood of the event. On the right, the event is assumed to happen, and the potential effects and the relevant post-hoc controls that could minimize their impact are listed \cite{koessler2023risk}.  

For example, given a hazard where an AI model has the capability of generating potentially harmful outputs, a risk event may be an AI providing information on how to create dangerous bioweapons. The risk factors could include the use of dangerous data for training as well as a lack of fine-tuning prior to deployment. The risk effects may include creation and use of bioweapons using the AI generated information. Once these risk factors and risk effects are in place, both preventive controls and post-hoc controls can be planned, such as appropriate filtering of training data and rigorous red teaming prior to model deployment as preventive barriers, as well as know-your-customer policies and model output censoring techniques as reactive barriers.

\rmm{System-theoretic process analysis (STPA)} STPA is a method to assess the utility of implemented controls against a particular risk within a complex system. Unlike bow-tie analysis, STPA factors in the interactions between components as events that can cause the risk in question \cite{koessler2023risk}. 

First, the system and its boundaries to the environment are defined. The system is primarily delineated from the environment because there is at least some partial control over it. Second, several items are enumerated, including (i) unwanted risk events (“losses”), (ii) system states that cause losses (“system-level hazards”), and (iii) system states that do not cause losses (“system-level constraints”). Third, a diagram mapping the system, environment, their different controls, and the interactions between these elements is created. This diagram must be comprehensive in listing the different losses and possible interactions that can cause each loss. Finally, the diagram can be used to identify “unsafe control actions” (UCAs), which are the causal pathways between a control and system-level hazards, including all interactions involved. 

For example, in the context of text-to-image models, losses may include ‘loss of diversity’ and ‘loss of quality’; hazards may include ‘low quality text-image pairs within training dataset’ and ‘harmful content within training dataset’; and the controls may include human controllers and automated controllers (e.g., annotators, data owners, data crawlers) \cite{rismani2023beyond}. Subsequent analysis may result in identification of UCAs such as ‘current data filtering actions’ and  neglecting current filter thresholds’ which are linked to specific hazards. Specific actions that counter or prevent such UCAs can reduce the associated losses.

\rmm{Risk matrices} A risk matrix is a method for risk evaluation. It is a heatmap that, for each cell, shows the severity score weighted by the likelihood score of a particular risk, usually from a scale of 1-5. Two rankings are required to construct a risk matrix: a ranking for the severity of risks, and a corresponding ranking of the likelihood of risks \cite{koessler2023risk}. 

AI-related risks can be generated using appropriate taxonomies, and placed into the relevant cells according to their assessed likelihood and severity based on predefined criteria (e.g., likelihood level 1 corresponds to $<1$\% chance, and likelihood level 5 corresponds to $>90$\% chance; while severity level 1 corresponds to mild inconveniences to the user, and severity level 5 corresponds to a fatality or financial damage upwards of \$10 million, etc.), such that particular focus can be given to mitigating risks with higher weighted scores (i.e., likelihood multiplied by severity). 

\rmm{Pre-allocate sufficient resources for risk management} The process of conducting thorough risk management is potentially time-\hspace{0pt}consuming. Pre-allocating sufficient resources, in terms of personnel count and schedule allowances, to conduct necessary risk management activities prior to model deployment is crucial \cite{anderljung2023frontier}. 

For example, a red-teaming exercise requires creative approaches to identify weaknesses of the AI system against potential adversaries, which alone may require hundreds of hours from several experts.

\subsection{Model Release}

\rmm{Staged release of model weights} When a model is developed by a provider for use in a certain AI system, it may also be useful to release the model itself more widely. Such developers can follow a staged release approach, in which they first grant access to the model via an API to trusted partners or the public, in order to scope the models’ capabilities and detect harmful or dangerous features \cite{solaiman2019release}.

After a period of an initial closed release and potentially further safety-training, the developers of the AI model can then release the weights, if they are confident that the AI model poses minimal systemic risk.

\rmm{Gradual or incremental monitored release of model access} AI systems can be released for access incrementally, starting with a small and selected deployer base before progressively being released to a wider user base. Initially, usage to a hosted API can be restricted with access given to specific deployers only, where all instances of the system can be easily updated or decommissioned with minimal disruption should there be any problems identified. Gradual releases provide more time to monitor for vulnerabilities and other problems. Even when such vulnerabilities are detected, the resulting harms may be more limited compared to a scenario in which a more capable version is released with the same vulnerabilities \cite{solaiman2019release}.

\rmm{Limit deployment scope} AI models can be restricted in terms of its use cases, where providers can require the deployers to limit its deployment to a predefined scope \cite{Github_2023}, such that models built for specific purposes and tested under specific environments are not used in environments or for purposes that are potentially unsafe.

\rmm{Restricted usage terms for open-source models} Developers of open-weights and open-source AI models can vet and restrict the users of their AI systems by requiring them to sign a Terms of Service agreement before getting access to the model weights. Such agreements can include limitations to the usage, modification, and proliferation of the AI model \cite{henderson2023self}.

Such agreements have the advantage that users only need to be vetted once before getting model access, but are often limited in practice in preventing unauthorised use or distribution. 

\rs{Non-decomissionability of models with open weights} If the model parameter weights are released or leaked in a security breach, the model cannot be decommissioned because the developer no longer has control over the publicly available model or its use. This prevents effective management and control of an open-sourced or leaked model. Models with publicly available weights are also easier to reconfigure, enabling misuse \cite{seger2023open}.

\rmm{Release strategy disagreement between developers} Developers of restricted-access models with similar capabilities may disagree about the strategy or precautions to take for model release, especially in the case of competitive pressure or minimal safety regulation oversight. In such a case, if only a single developer releases an equally capable model unrestricted, malicious actors can use it instead of restricted-access alternatives \cite{henderson2023self}.

\subsection{Post-Deployment Practices}
\rmm{Pop-up interventions in LLMs} Supplementary information can be shown to the user in specific query topics where factual accuracy is critical. This intervention can effectively divert users from potential inaccuracies generated by AI models in sensitive contexts. For example, during electoral processes, where model hallucinations can be particularly costly or have a negative impact on society, LLMs can offer their users the option of being redirected to accurate and up-to-date information sources \cite{Anthropic_2024_elections}.

Since pop-up interventions can be intrusive to workflows, they are best used in situations where the benefits of the information outweigh the cost of distraction.

\rmm{AI identification} AI identifiers can be used to indicate that an AI is involved in a process or an interaction \cite{chan2024visibility}.

For AI systems that interact directly with users, a visible output may be used, e.g., a displayed text message saying “I am an AI language model”, accompanied with the appropriate warnings and caveats relevant to the user, \cite{chan2024visibility}. Whereas, for AI systems that interact with other systems or applications, other forms of watermark or unique identifiers can be used. In either situation, agent cards can serve as an identifier, where further details about the underlying AI system, the specific instance of the AI agent, and other information relevant to the development of the agent, can be included. 

\rmm{AI output watermaking} Output produced by or with AI assistance can be marked to clearly identify its origin. Verification of the watermark can involve the use of statistical tests or having the mark immediately visible to a human inspector. Ideally, the watermark does not significantly alter the utility of the output, and is robust against digital and physical manipulation that results in data degradation \cite{werder2022establishing, OpenAI_metadata}.
 
\rmm{AI output metadata} Output produced or whose production is aided by AI can contain metadata to record its origin and the transformations it has undergone. Metadata is evidence that subsequent versions or its derivatives come from this original version. The metadata can include the original AI model source, along with ownership, and its subsequent edits \cite{rosenthol2022c2pa}. 

For example, an image produced by an AI model can contain metadata showing the date of creation and the AI model that produced it. Subsequent versions can reference this information and, if they are intermediary versions, can include descriptions of any editing that has taken place. 

\subsection{Monitoring}
\rmm{Monitoring of model capabilities} AI models are often trained to develop specific capabilities by using appropriate training data and training goals. However, models may develop capabilities that they were not specifically trained for. One subset of this is emergent capabilities, i.e., capabilities that emerged in larger models but not smaller models given a similar training process \cite{wei2022emergent}.

These capabilities can be monitored, allowing models to be tested not only for their intended capabilities but also for capabilities that are not intended.

\rmmEX{AI model-assisted oversight of AI systems} AI model-assisted oversight can help monitor and supervise the training of increasingly capable GPAI systems, which may become difficult to oversee at scale by human supervisors during training or testing. Monitoring and supervision may become especially difficult in cases where increasingly advanced GPAIs perform near or above human level in some specialized domains, where supervision quality might fail to keep pace with capabilities improvement. The training signal may include labeled data, reward function, and user feedback on produced outputs. 

Currently, there are two broad approaches to provide scalable training signals to such systems:
\begin{enumerate}
    \item \emph{Scalable oversight}: Improving  of the supervisor’s capabilities to supervise, such that they can provide accurate training signals quickly and at scale \cite{bowman2022measuring}. 
    
    For example, a debate format can be used between two GPAI systems (two instances of the same GPAI, or similarly capable systems). A Human supervisor judges the debate, making it easier to assess correct responses in domains which might otherwise require significant time investment of domain specific expertise \cite{khan2024debating}.

    \item \emph{Weak-to-strong generalization}: Enhance the training signals while ensuring that the enhanced signal remains faithful to the intentions of the original human-provided signal \cite{burns2023weak}. 
    
    For example, a hierarchical (“bootstrapping”) oversight approach can be implemented: A series of GPAI models with increasing capabilities are used, where each model in the hierarchy provides oversight for the next more capable model. The least capable model at the base of the hierarchy is the only one directly overseen by human supervisors, as it is easier to oversee than the more capable models.
\end{enumerate}

\rmm{Post-deployment ongoing monitoring for unexpected use cases} Ongoing monitoring of AI systems can uncover emergent or newly identified capabilities and limitations, in particular when new use cases are found, or in a large-scale deployment with a diverse population of users. These new capabilities or limitations can inform ongoing risk analysis. New use-cases can be discovered via monitoring publications, online forums, or APIs \cite{matsui2022mlops}. 

For example, a LLM might have unexpectedly high competence at giving convincing medical advice, despite not being directly developed for that purpose, nor verified for accuracy. In that case, the potential risks of this newly found competence can be assessed. 

\rmmEX{Monitoring protocols for potentially adversarial GPAI systems} Similar to cybersecurity protocols, AI providers can prepare in advance for potential adversarial action from AI systems by developing and stress-testing protocols for potential adversarial action \cite{Sehgal_2018}. Such adversarial action may range from hiding code exploits, to strategic deception or loss of control related behaviors. 

AI systems may be adversarial due to malicious actors (e.g., data poisoning), or due to unintended issues of control (lack of alignment with human intent). Alongside improving the system alignment, external monitoring and restrictions may be deployed in advance to “catch” these adversarial model malfunctions before harm is done. 

Like with cybersecurity red teaming, any protocol which assumes adversarial action can be stress tested in advance. For example, this can be done by projecting the plausible capabilities and affordances of the adversarial actor (e.g., in the case of a tech company -  a hacker, or a disgruntled employee). From this process, measures that better anticipate possible adversarial action can be deployed, and catch unsuccessful attempts at adversarial behavior. 

If such unsuccessful attempts are caught, they can be reported in incident reports, and serve as an input to further risk analysis. Depending on their severity, it may be advisable to recall the system if deployed. 

Examples of these protocols include \cite{greenblatt2023ai}: 
\begin{itemize}
    \item Flagging of suspicious frontier AI system code generation for human oversight, with the help of better proven (“trusted”) AI systems or automated keyword detection. 
    \item Using a better proven (“trusted”) AI system to edit the code produced by a frontier AI system if suspicious.
    \item Monitoring of AI system with another instance of this same system, with filters to prevent collusion.
    \item Restriction of access of the AI system to private information. 
\end{itemize}

\rmm{Encouraging downstream provider to evaluate models for deployment-specific failure modes} In some cases, AI system deployers are better positioned to perform certain risk management measures on the AI model in a provided AI system, relative to upstream model providers. For example, they understand their use case better and are more easily able to predict foreseeable misuse or failure modes. These evaluations can inform upstream model providers, or inform supplementary mitigations by the deployer.

\rmm{Encourage reporting of critical vulnerabilities to the upstream provider or other relevant stakeholders} Downstream AI system deployers can report critical vulnerabilities or incidents to the upstream model provider and other relevant regulators. This can contribute to safe use, and allow other downstream deployers to be informed about any potential problems.

\section{Cybersecurity}
\label{Cybersecurity}

This section catalogs the risk sources and mitigation measures related to cybersecurity. These items may be related to security in terms of AI models being accessible only to the intended users, as well as AI models having appropriate access to the external world during both model development and deployment stages.

\rs{Interconnectivity with malicious external tools} The growing integration and interconnectivity with external tools and plugins increase the risk of exposure to malicious external inputs. This interconnectivity makes it easier for external tools to introduce harmful content \cite{xie2023adaptive}.

\rs{Unintended outbound communication by AI systems} AI systems that have the broad ability to connect to a network to obtain information could also end up sending data outbound in ways that neither providers, deployers, or end users intended \cite{nakano2021webgpt}. This can happen if there is no whitelisting of communication channels (such as network connections or allowed protocols). In general, this can occur if the deployment of the AI system violates the principle of least privilege.

Such outbound communication may lead to leakage of confidential data, or the AI system performing unwanted actions like sending emails or ordering goods on the internet.

\rs{AI System bypassing a sandbox environment} An AI system may have the ability to bypass a sandboxed environment in which it is trained or evaluated. 

For example, the AI system can achieve this by finding and using misconfigurations or vulnerabilities in the software of the sandboxed environment. This can also occur if the AI system finds and uses vulnerabilities of the hardware it is being run on, or by using social engineering techniques on the users or administrators of the sandboxed environment \cite{fang2024llm}. 

The developers or malicious actors may intentionally create such behavior (e.g., by inserting backdoors), or it can occur unintentionally, with the AI system bypassing the developer-intended domain of operation \cite{achiam2023gpt}.

\rmm{Least Privilege access} Deployers of an AI system can restrict its permissions to a whitelisted set of predetermined options, such that all options not on the whitelist are not accessible to the AI \cite{tang2024prioritizing, OWASP_2023}.

The entries on the whitelist can be chosen to be as small as possible for the AI system to fulfill its intended purpose, to reduce the attack surface of external attackers, and to decrease the probability that the AI system accidentally takes actions with large unintended side-effects.

For example, such whitelisting can apply to network connections, execution of other programs, access to knowledge bases, and the action space of the AI system. It can be implemented through running the AI system on an OS-level virtualization, on networks behind a firewall, and in extreme cases on air-gapped machines.

\rs{Model weight leak} Model weights or access to them can be leaked when initial access is granted only to a select group of individuals, such as institutional researchers \cite{Vincent_2023}. This risk can increase as more people gain access, and identifying the source of the leak becomes more difficult. The availability of leaked model weights makes various attacks on systems that use the leaked AI model easier to implement, such as finding adversarial examples, elicitation of dangerous capabilities, and extraction of confidential information present in the training data. The availability of model weights might also enable the misuse of the AI system using the leaked model to produce harmful or illegal content \cite{eiras2024near}.

Model weights are typically controlled by the developers, who may grant access to authorized parties for specific purposes (e.g., for red teaming, researcher access, and early access to downstream deployers for ``beta testing"). However, if these authorized individuals share the weights with others without permission, it creates unauthorized access. This makes it challenging to maintain control over who has access to the model weights.

\rmm{Protect proprietary or unreleased AI model architecture and parameters} The developers of AI models can invest in cybersecurity to prevent compute resources, training source code, model weights, and other critical resources from being accessed and copied by unauthorized third parties (e.g., through insider threats or supply chain attacks). 

Access to model source code and weights can be restricted through an access control scheme, such as role-based access control. If access to model outputs by third parties is required, it can be provided through an API. Air gaps can block unauthorized remote access. In the case of necessary interaction with an external network, network bandwidth limitations can also be enforced to increase the detection window of potential breaches \cite{koessler2024risk}.

\rmm{Hardware limitations on data center network connections} Hardware-enforced bandwidth limitations on data center network connections can protect AI model weights from unauthorized access or exfiltration, by limiting the speed of model weight access on the connections between data centers and the outside world.

Such limitations can be put in place in multiple ways, for example by only constructing connections with a specific bandwidth. The output rate on all data channels can be set low enough that copying the weights is possible in principle (e.g., to enable regular backups), but would take so long that an unauthorized exfiltration of the weights could be detected and prevented.

Such rate-limiting is only effective if it applies to all output connections for all storage locations on which the weights of the model are stored \cite{nevo2024securing}.

\rmm{Structured access to a model} Structured access refers to methods which limit users' or deployers' direct access to a model's parameters by constraining access to a model through a centralized access point (e.g., an API) \cite{henderson2023self}. This access point can be monitored for usage, and access can be revoked to users or downstream deployers in cases of misuse \cite{kim2024jailbreaking}. 

Within this centralized access point, automated filtering-based monitoring can be done on both inputs and outputs to ensure the model’s intended use is preserved \cite{bucknall2023structured}. This filtering can sometimes be supplemented by human oversight, depending on desired robustness levels. 

\rmm{Sandboxing of AI Systems} AI systems can be developed and tested within a sandbox, (a secure and isolated environment used for separating running programs), such that outside access to information within the sandbox is restricted. Within this environment, resources such as storage and memory space, and network access, would be disallowed or heavily restricted \cite{babcock2019guidelines}. With sandboxing, dangerous or harmful outputs generated during testing will be contained. 

\section{Impacts of AI}
\label{Impacts}

This section catalogs the risk sources and risk management measures related to the impacts of AI systems. These impacts may be direct, resulting from the use of an AI system (e.g., physical, societal, or financial impacts), or they may be indirect, unrelated to the direct usage of model outputs, such as the use of data (e.g., privacy concerns) or the energy consumption during model training and inference (e.g., environmental impact).

\subsection{General}
\rs{High-impact misuses and abuses beyond original purpose} Since general-purpose AI systems have a large repertoire of capabilities, malicious actors such as foreign actors can use such systems to cause large damage if they gain unrestricted or unmonitored access to those AI systems.

For example, LLMs have strong source code generation capabilities, which can also be used to aid the creation of malware or the discovery of exploits in existing software and hardware.

Relevant AI systems are ones that can help to create or can be used as weapons, such as lethal autonomous weapons (through object and movement detection and maneuvering), bioweapons (by enabling non-specialists to perform specialist lab work) \cite{urbina2022dual}, cyberweapons (through finding and creating exploits), and other military technologies.

\rs{Democratizing access to dual-use technologies} Access to dual-use technologies can become easier because of GPAI model proliferation (in particular, open-source or open-weights models). Non-experts can use such dual-use-capable systems at a minimal cost \cite{soice2023can, kang2024exploiting}. Improved model capabilities also contribute to dual-use risks posed by malicious actors.

For example, an open-source base model for generating high quality sequence data can be modified to generate candidate protein sequences for toxin synthesis \cite{boiko2023emergent}.

\rs{Competitive pressures in GPAI product release} In competitive situations, developers of general-purpose AI systems might cut corners on the safety evaluation of their GPAI model and instead spend more time and effort on the capabilities of those systems \cite{shevlane2023model, emery2023uncertainty}. This is especially dangerous if the capabilities of such AI systems are correlated with the risk they pose \cite{ren2024safetywashing}.

For example, competitive pressures can be exacerbated by market competition, where GPAI providers are primarily developing products to sell. Given the prohibitive cost to develop large models, losing such competition can compromise companies financially. This situation can incentivize companies to prioritize financial survival over safety.

\subsection{Physical Impacts}
\rs{Damage to critical infrastructure} The integration of AI systems within critical infrastructure, ranging from transportation to power systems, can cause substantial damage in cases of failure or malfunction. With the increasing number of Internet of Things (IoT) devices and interconnected cyber-physical systems, critical infrastructure becomes even more vulnerable \cite{sakhnini2020ai, sayler2020artificial}. 

\rs{AI-based tools attacking critical infrastructure} Critical infrastructure can also be damaged without AI integration, for instance, when AI-based tools are used indirectly to aid actions such as in coordinated power outages caused by large-scale user manipulation \cite{raman2020weaponizing}.

\rs{Critical infrastructure component failures when integrated with AI systems} When relying on GPAI in critical infrastructure, there may be common mode failures that begin with vulnerabilities or robustness issues in the underlying model architecture or training setup. These failures may happen accidentally (in edge-cases) or due to adversarial inputs to the AI systems \cite{DHS_report_2024}.

For example, a failure in the scheduling software of a chemical plant caused by an adversarial keyword can cause damage to physical property through halting critical processes (e.g., cooling, mixing of reactants).

\rs{AI Systems interacting with brittle environments} Deployed AI systems can rely on physical sensors and data sources that may exhibit hardware drift and thus data distribution drift over time. This distribution drift may affect system robustness and performance. This usually involves AI systems working in undigitized and physical environments.

For example, for a traffic violation detection system, a slight camera movement due to environmental conditions can cause failures in detection \cite{sambasivan2021everyone}.

\rmm{Redundant systems not reliant on GPAI} Redundant systems provide continuation of a given system’s processes in case of failure of the given system. Importantly, redundant systems should not rely on the factors that caused the original system to fail in the first place, which can include an AI system \cite{conrad2012cissp}.

For example, if an AI system is incorporated into the landing gear system of an aircraft, such as during autonomous control of the aircraft, redundant systems in the form of mechanical or hydraulic mechanisms must be present to allow for deployment of the landing gear in case of AI system failure.

\subsection{Societal Impacts}

\rs{AI-generated advice influencing user moral judgment} AIs can easily give moral advice even when not having a coherent, contradictions-free moral stance. This could lead to the users’ moral judgments being negatively influenced by random or arbitrary moral advice given by AIs \cite{krugel2023chatgpt}.

\rs{Overreliance on AI system undermining user autonomy} AI systems can undermine human autonomy, if they allow for habitually trusting the AI’s suggestions without sufficient exercising of human agency. Over time, a user may develop unjustified trust in or dependence on the system, or rely on its advice for tasks outside the system’s domain of expertise \cite{vasconcelos2023explanations, chiesurin2023dangers}. In particular, less confident users (or users in emotional distress) can be more prone to “overtrust” a system \cite{Xiang}.

\rs{Automatically generating disinformation at scale} Disinformation (in various modalities: text, audio, images, video, etc.) can be generated with minimal human oversight and effort. Disinformation tools are relatively cheap and their technology is widely available. Such deployments can be particularly widespread in sensitive political contexts.

\rs{AI-driven highly personalized advertisement} Advanced GPAI systems can create advertisements tailored to individual recipients, exploiting the biases and irrational beliefs of each recipient. Such advertisements can cause consumers to make decisions they regret in retrospect, or would regret upon more reflection.

Current versions of personalized video advertisements already show better results compared to regular advertisements \cite{kumar2023generative}. However, the widespread use of highly personalized advertisements raises concerns about undermining consumer autonomy and exacerbating social inequality.

\rs{Generative AI use in political influence campaigns} GPAI tools can be used in automation and scaling of influence campaigns \cite{seger2023open}. Public opinion may be manipulated by targeted misleading or manipulative information. This can lead to rising political polarization and diminishing trust in public institutions.

\rs{Generation of illegal or harmful content} Generative models can create illegal, harmful, or discriminatory content \cite{solaiman2023evaluating}, such as sexual abuse material, at scale. Current access controls (e.g., API access filters) are not effective against all user queries in generating such content.

\rs{Unintentional generation of harmful content} Generative models can create harmful or discriminatory content from benign user requests. Models can exhibit bias to particular harmful styles of generation (e.g., sexualization of photos of women \cite{technologyreviewViralAvatar} in the case of image generation models) or they can generate toxic, misleading, or violent data (e.g., a model generating jokes can use ethnic stereotypes or slurs to deliver humor).

\rs{Multimodal deepfakes} Deepfakes are media that depict real or non-existent people or events, involving the use of multiple modalities (e.g., images, audio, video). They can also involve the imitation of speech or body movements of real people. Multimodal deepfakes can be used to harass, discredit, intimidate, and extort individuals.

\rs{Generation of personalized content for harassment, extortion, or intimidation} GPAIs can be misused for the automated generation of content personalized to target select individuals based on their weak spots \cite{bommasani2021opportunities}. Such attacks may be more efficient and more successful in achieving the goals of harassment, extortion, or intimidation.

\rs{Misuse for surveillance and population control} AI tools can be misused by human or institutional actors for monitoring, controlling, or suppressing individuals \cite{seger2023open}. Massive data collection and automated analysis are often conducted, and AI tools can further exacerbate such practices.

\rs{Systemic large-scale manipulation} AI systems embedded with systemic biases can manipulate large population segments, particularly when these biases align with the beliefs or behaviors of the targeted group. When weaponized at scale, this manipulation can exacerbate social divisions or cause large-scale disruptions, such as city-wide blackouts (e.g., by the manipulation of power consumption into the peak demand period \cite{raman2020weaponizing}).

\rs{Diminishing societal trust due to disinformation or manipulation} The use of GPAIs may contribute to the proliferation of either deliberate disinformation or unintended misinformation can severely erode trust in public figures and democratic institutions. This diminishing trust can extend to other forms of media, making the public less informed.

\rs{Personalized disinformation} Automatic generation of disinformation can be personalized to target specific groups or individuals. Such attacks can be more effective in achieving their goals, and their costs can be significantly reduced when using GPAIs.

\rs{GPAI assisted impersonation} GPAI outputs are not always correctly detected as AI-generated across multiple modalities (text, images, audio, video). A malicious actor can use GPAI outputs directly when communicating, or use AI-informed details to help construct a convincing impersonation (e.g., forging of supporting documents). 

Even if future countermeasures prove potent enough to detect GPAI-generated content, the risk remains if the countermeasures are not well known, or difficult to access.

\subsection{Financial Impacts}

\rs{Deployment of GPAI agents in finance} The deployment of GPAI based agents in the financial sector can negatively impact market stability due to correlated autonomous actions, high interconnectedness, or incentive misalignment \cite{aldasoro2024intelligent}. Furthermore, such GPAI agents in the same environment are vulnerable to classical challenges in multi-agent systems \cite{dorri2018multi}, such as coordination and security of the agents.

For example, an agent tasked with predicting the value of a commodity, for which numerous agents depend on to make their own predictions, can be targeted with unreliable data, compromising the actions of both the main agent and its dependents.

\rs{Financial instability due to model homogeneity} The widespread use of similar models or algorithms across the financial sector can lead to synchronized reactions to market signals, increasing volatility, triggering flash crashes, or market illiquidity \cite{aldasoro2024intelligent}.

\rs{Use of alternative financial data via AI} Alternative financial data of a company is any data about the company not produced by that company. Examples of such data that can benefit from improved collection and aggregation using AI models include stock discussions on social media, product reviews, and satellite imagery.

The use of alternative financial data, enabled by the deployment of AI models, may introduce biases and generalization issues due to shorter shelf-life and varying quality (e.g., shorter time series, smaller sample sizes, and dubious claims) due to its origins from various sources, posing financial tail risks (i.e., tail-end of a probability distribution), where the price of a company changes dramatically \cite{aldasoro2024intelligent}.

\subsection{Cyberattacks} 

\rs{Automated discovery and exploitation of software systems} GPAIs can be used to aid in the automated discovery of software vulnerabilities \cite{brooks2019survey}. This can empower malicious actors, making their cyberattacks more efficient and potentially more damaging. This type of automation allows attackers to expand the scale of their operations at a low cost, increasing the impact of their actions. New malware can be developed automatically, or the known vulnerabilities can be exploited to create more sophisticated attacks.

\rs{Amplification of cyberattacks} General-purpose AI models may significantly enhance the magnitude and effectiveness of cyberattacks, by amplifying existing capabilities or resources of malicious actors \cite{aisiAdvancedEvaluations}. For example, GPAI models may be employed to:

\begin{itemize}
    \item Automatically scan open-source codebases and compiled binaries for potential vulnerabilities
    \item Apply known exploits flexibly and at scale (e.g., identifying vulnerable computers based on subtle cues in response times or output formats)
    \item Assist with different aspects of cyberattacks, including planning, reconnaissance, exploit searching, remote control, malware implementation, and data exfiltration
    \item Combine social engineering (phishing, deepfakes, etc.) with cyberattacks at scale. 
\end{itemize}

\rs{AI-driven spear phishing attacks)} Generative models can be misused to target individual users more efficiently by using personalized information \cite{barrett2023identifying}. Highly convincing automated fraudulent schemes can exploit the trust of victims by extracting sensitive data and making the deception more likely to succeed. For example, in LLMs, this misuse can be aided by jailbreaking techniques \cite{seger2023open}.

\rs{Models generating code with security vulnerabilities} Models can generate code or coding suggestions that contain security vulnerabilities. This may occur across various LLM-based model families, including more advanced models with superior coding performance, where the tendency to produce insecure code is even more pronounced \cite{bhatt2023purple}.

\subsection{Weapons} 

\rs{Misuse of AI systems to assist in the creation of weapons} AI systems may be misused to aid in the creation of weapons, such as chemical, biological, radiological, and nuclear (CBRN) weapons, or augment the abilities of existing weapons, such as providing autonomous capabilities to unmanned weapon systems. Current systems do not significantly aid a malicious actor in these tasks, but they do show early signs \cite{li2024wmdp}. This risk can sometimes be mitigated with input and output filtering, but is still susceptible to adversarial techniques (such as jailbreaking or paraphrasing).

Examples of tasks which GPAIs may help perform include:

\begin{itemize}
    \item Creation or implementation of plans. For example, this may involve route planning for drones or generating weapon component schematic diagrams for weapon creations.
    \item Technical assistance R\&D or manufacturing. For example, this may involve helping troubleshoot a chemical process.
    \item Simulation or code scripts. For example, this may involve helping interface with more complex simulation software (e.g., fluid dynamics), or producing analysis more quickly with less advanced simulation software specific expertise needed compared to using simulation software directly.

\end{itemize}

\rs{Misuse of drug-discovery models} Models used for drug discovery, such as drug-target affinity prediction models, can be used to identify or develop dangerous toxins. This is particularly concerning if the training data contains information related to potentially dangerous proteins and viruses.

\subsection{Bias} 

\rs{Homogenization or correlated failures in model derivatives} Homogenization refers to common methodologies and models used across downstream GPAI systems, which may lead to uniform failures and amplification of biases \cite{schneider2024foundation, bommasani2021opportunities}. This risk arises when numerous downstream AI systems are built upon a few large-scale foundation models. 

This may be caused by centralization of AI advancements within a few companies, as well as flaws from algorithmic monoculture, where dataset sources and collection methods are similar across numerous AI models.

Homogenization in models may lead to consistent and arbitrary rejection, mistreatment, scrutiny, or misclassification of specific users of groups, as well as the spread of implicit perspectives (e.g. bias towards a particular political group) across multiple application domains.

\rs{Reporting of user-preferred answers instead of correct answers} AI systems with natural-language outputs can tend to give answers that appear plausible or that users prefer \cite{perez2022discovering} but are factually incorrect. This phenomenon is sometimes referred to as “sycophancy.”

This behavior can occur if the AI system is updated after human users give feedback on the outputs of the model, since human feedback has systematic biases which an AI model can learn from. In such a case, the reinforced behavior can favor giving inaccurate but human-preferred answers, where the preference is inferred from cues in the input.

AI models giving preferred but incorrect answers can also happen if they are configured by model developers to do so, in order to make the resulting product more palpable to consumers. This can also happen if they are trained on data which contain many conversations between people who agree with each other.

\rs{Biases in AI-based content moderation algorithms} AI-based content moderation algorithms, while intended to filter harmful content, can perpetuate biases. For example, gender biases within these systems may lead to the disproportionate suppression or “shadowbanning” of content featuring women \cite{theguardianThereStandard}. 

AI moderation tools may embed and reinforce the objectification of women by classifying and rating images of women as more sexually suggestive compared to similar images of men \cite{theguardianThereStandard}. This can result in the unintended marginalization of female-led businesses and contribute to broader societal inequalities.

\rmm{Diverse data labeling and algorithm fairness audits} To mitigate biases in AI models, model providers may want to prioritize diversity among data labelers and conduct regular fairness audits on their algorithms. Data labeling teams that represent different backgrounds and demographic groups can help create more balanced datasets.

\rs{Systemic bias across specific communities} AI systems may exhibit unfair or unfavorable outputs across a range of tasks against specific communities of people, either implicitly or explicitly. Bias can lead to forms of exclusion or erasure (e.g., mislabelling for categorization-based tasks) and violence (e.g., sexual violence against women from deepfake pornography).

These biases are systemic because they come from both technical and non-technical factors affecting the development of the model. Relevant factors include the training data, the system’s intended use and design, and its governance structure that can exclude accountability on affected issues. 

Such biases can mutually reinforce each other as AI systems become entrenched into the socio-political environment of these communities \cite{ashwini2024contemporary}, especially when biased outputs become inputs of other AI systems.

\rs{Unintentional bias amplification} Dataset bias may be unintentionally amplified \cite{dinan2019queens} where the outputs of the AI model trained on a dataset are more biased than the dataset itself.

\rmm{Debiasing methods} Providers of AI models can apply techniques to reduce the biases of their models.

Current debiasing methods focus on three main types of bias:

\begin{itemize}
    \item Racial and religious bias - Stereotypes based on religious beliefs or racial beliefs.
    \item Gender bias - Stereotypes tied to gender roles and expectations.
    \item Political and cultural bias - Propagation of dominant ideologies or extremist attitudes.
\end{itemize}
    
Debiasing methods can be categorized based on their application during AI development:

\begin{itemize}
    \item Data pre-processing - Removing or correcting unwanted and biased data, and augmenting quality data to offset data bias, such as rebalancing datasets with counterfactual data augmentation.
    \item During training - Intervening on the training dynamics of the AI model, such as introducing debiasing terms in the objective function or by negatively reinforcing biased outputs.
    \item Post-training - Applying techniques to correct a trained but biased model, such as modifying the embedding space.
\end{itemize}

\rs{Long-term effects of AI model biases on user judgment} The initial user exposure to model biases can have a lasting impact beyond the initial interaction with the model. Users who encounter biases in AI models can be affected by and continue to exhibit previously encountered biases in their decision-making, even after they stop using the models \cite{vicente2023humans}.

\subsection{Privacy}

\rmmEX{Knowledge unlearning techniques} Knowledge unlearning techniques allow specific information to be “forgotten” without the need for retraining the entire model, preserving its general capabilities. These techniques can be used to reduce privacy risks and protect against copyrighted or harmful content \cite{jang2022knowledge, si2023knowledge}.

\rmm{Differential privacy} Differential privacy techniques \cite{anil2021large} can be used to protect users’ privacy by ensuring that sensitive information is not leaked from a training dataset, even after thorough statistical analysis. With differential privacy, noise is added to the dataset or the model’s output in such a way that one cannot deduce the presence or absence of a particular data point within the dataset. This provides individuals with plausible deniability and prevents their information from being exposed.

\rmm{Quantifying privacy risks of AI models} Measuring privacy risks of an AI model allows the provider and user to calibrate their expectations on where the model can be applied, and it allows them to take the necessary steps to reduce such risks. 

For example, some metrics include:

\begin{itemize}
    \item Success rate of membership inference attacks \cite{shokri2017membership}  -  Measures the rate that an attack correctly predicts a given record is part of the training dataset used to train a given AI model. 
    \item Discoverable memorization \cite{carlini2022quantifying} - Theoretical upper-bound of the amount of training data that a given model memorizes. Assuming full knowledge of the training data, it measures the percentage of items that, for a given incomplete data point, a model outputs the remaining (memorized) part.
\end{itemize}

\rs{Decision-making on inferred private data} Current GPAIs (LLMs and multimodal LLM-based models) have significant capability to infer correlations in text data. In some cases, they may be able to make highly accurate data inferences on users based on contextual input that users provide \cite{meisenbacher2024privacy}. These data inferences can “leak” or reveal sensitive information about the user, cause unfair treatment, or enable manipulation of user behavior. 

Some information that can be inferred from user input may include age, gender, political leanings, and country of birth. While this information might not be present explicitly in the data, it may be easier for a GPAI system to infer this information compared to a human. 

This capability may be used for both intentional manipulation (e.g., personalized or targeted advertising, malicious actors using GPAIs for influence campaigns) or unintentional manipulation (e.g., different responses to factual questions by models trained to be agreeable or helpful, when asked by different demographics). 

\subsection{Environment}

\rs{High energy consumption of large models} Training and deploying large models require substantial energy expenditure. The trend toward developing larger models exacerbates this issue. This can lead to excessive energy usage and have a negative environmental impact. 

\rmm{More energy-efficient models or techniques} Deploying more energy-efficient models can reduce their environmental impact. Different model architectural choices result in varying environmental costs, and identifying and adopting more energy-efficient options can result in significant environmental savings, especially when implemented at scale. Consideration should be given to both training energy usage, and deployment (“inference”) usage for the expected model lifecycle.

\rmm{Disclosure of energy consumption by AI systems to authorities} Disclosures can direct more necessary attention and scrutiny to projects that consume significant energy. Disclosure involves releasing a summary of key details of the energy consumption of the AI system by all users, including the compute resources used, the amount of power consumed, the measures to reduce excess energy consumption that were in place, and energy sources \cite{mcdonald2022great}.

\rmm{Using low carbon intensity energy grids} Moving model training to energy grids with low carbon intensity can reduce the negative environmental impact \cite{bommasani2021opportunities}. The efficiency of energy grids can vary greatly depending on location. Models can be trained in different locations, as latency is not an issue.

\section{Discussion}
\label{sec:disc}
In this section, we discuss various insights about the different risk items we have cataloged. We present important challenges in risk assessment, which mainly focuses on identifying risk sources and harms, as well as important challenges in risk management, which includes risk management measures and internal provider governance. Finally, we consider the role of small-medium enterprise (SME) GPAI providers, including their unique challenges and considerations for them.

\subsection{Challenges in Risk Assessment}

We believe some of the greatest challenges in risk assessment arise as a result of false negatives: when providers mistakenly conclude that they have sufficiently managed and evaluated systemic risk. For example, red teaming or evaluation-related efforts are necessary but not sufficient. Even if red teaming or benchmarking efforts do not elicit dangerous behavior patterns or expose a lack of robustness, it does not guarantee that the model is safe to deploy.

Evaluation processes, in addition to red-teaming, are helpful in assessing risk level of AI systems both pre and post deployment. However, these evaluations (automated or with human oversight) may fail to catch the risks stemming from the AI system if their scope is too narrow or limited and lead to costly halting of later stages of development. This is akin to a failure of quality control in one part of the assembly line. Further discussion on benchmarking is in Appendix \ref{sec:benchmarking}.

\subsection{Challenges in Risk Management}

We believe that in the quickly advancing field of GPAI, some of the greatest challenges in AI risk management are allocating the necessary resources for risk management measures and dealing with uncertainty within internal organization decisions.

In many cases, proper implementation of risk management measures may require significant time and effort that may compete with the other organization-level obligations of the provider, such as product ship dates. For example, a red-teaming exercise may take hundreds of hours with the involvement of multiple 3rd party experts with diverse knowledge \cite{anthropicFrontierThreats}, where issues uncovered may also delay a product launch. A significant fraction of the providers' personnel may be needed to organize regular exercises, document findings and turn them into actionable insights, and coordinate red teaming participants within multidisciplinary teams. The costly requirements for red-teaming exercises can go against the interest of the provider to remain economically competitive. There is then a persistent temptation to re-allocate staff to develop their GPAI into a viable product instead.  At worst, this can develop into a form of “safetywashing" \cite{ren2024safetywashing}, where the surface-level or misleading impression of safety is prioritized over actual risk management efforts (in a way analogous to “greenwashing").

For internal governance within an AI provider organization, a recurring challenge will be making decisions about risk under uncertainty. This is due to the difficulties in defining and measuring risk thresholds in these relatively new and untested GPAI systems \cite{koessler2024risk}. In this situation, it may be useful to remain very agile, in response to novel risks arising during model development,  while also having rigorous internal practices and risk culture to keep track and respond to risks as they emerge. 

Realistically, a provider will face challenges in achieving the desired outcomes even with implementation of relevant risk mitigation measures. Some examples identified in the literature include:

\begin{itemize}
    \item Scalability issues: risk mitigations may not generalize to future GPAIs with more advanced capabilities \cite{debenedetti2023scaling}.
    \item False sense of safety, e.g., models that are deliberately trained to bypass safety training may appear to be safe \cite{hubinger2024sleeper}.
    \item Limited understanding of evaluations and safety benchmarking \cite{ren2024safetywashing, zhang2023safetybench}.
\end{itemize}

Given these challenges, GPAI providers should be prepared to develop (or wait for the development of) further safety-related innovation and have procedures in place for deciding whether to stop the deployment of a model if it fails fundamental risk assessments, even in the face of competing interests.

\subsection{Risk Management for Small-Medium Enterprise (SME) GPAI Providers }

We envisage that risk management will broadly apply to different types of GPAI providers, including those that have limited size and capacity, such as SMEs. SME providers may make very different tradeoffs in managing the risks due to the dual-use nature of a GPAI model. One provider may severely restrict and control potential GPAI model users, while another may go full open source while severely restricting dangerous model capabilities or capabilities leading to that. We believe that regulations should account and allow for this type of flexibility.

Note also that there are GPAI models that can be considered to not have systemic risks and/or have a very limited and controlled user base. Therefore, not all of the measures described are necessary or even appropriate, as the risk reduction from implementing such measures may be minimal. Especially for SMEs and startups developing GPAI models \emph{found to not have systemic risk}, minimal reductions in risk may not justify the costs incurred. An example is the cost of hiring third parties to implement some risk management measures.

Finally, for provisioning tools for risk management in general, regulation should further envisage and promote the creation of a market that can help providers of variable size access to the tools and multidisciplinary capabilities needed to perform state-of-the-art risk mitigation. One example may be provisioning such tools in the form of public goods, where a suite of tests and infrastructure are hosted by a public institution and are freely available for organizations and individuals to evaluate the safety of their models. 

\section{Conclusion}

We introduce a risk catalog broadly covering risk sources and risk management measures relevant to GPAI systems that have systemic risk. Our catalog covers the entire model development lifecycle, from design and training to evaluation, and deployment. We also look beyond the GPAI model itself and analyze risks relevant to fundamental components of GPAI development such as data, deployment infrastructure, users, and governance practices by providers. We also cover impacts of the use of GPAI systems along multiple dimensions ranging from social, environmental, physical, military, and economic impacts. Finally, we discuss various considerations relevant to both GPAI providers and regulatory frameworks in using the items in our risk catalog.

We hope that our work can contribute to the development of standards and further governance efforts by providing balanced and accessible descriptions of key GPAI-related risks and their mitigation measures, as well as drawing attention to those associated with lesser-known risks.

\section*{Acknowledgments}
We thank Giles Edkins, Kathrin Guardhouse, David Manheim, Sofia ``Zow'' Ormazabal, Chun-Wei Tsai, and Yonatan Cale for providing extensive feedback on an early draft of this paper. The content of this paper represents the views of the authors and does not necessarily reflect the views of the people listed here.

\newpage
\bibliography{main.bib}
\bibliographystyle{acm}

\newpage
\startappendix

\section*{Appendices}
\addcontentsline{toc}{section}{Appendices}

\section{Further Information on GPAI Technology and Value Chain}
\label{sec:gpai_tech}
In several ways, GPAI technology represents a break from existing design approaches, which we may also call paradigms, for how machine learning is used to construct useful AI systems.

\begin{enumerate}
    \item An existing design approach for constructing a useful and robust predictive AI model, one that can predict a future outcome $Y$ based on current observations $X$, is to compile as many accurate historical $(X, Y)$ pairs as possible, and train a model using that dataset.
    
    \item An existing design approach for constructing a useful and robust AI agent is to configure a machine learning algorithm with a reward function, and then have that algorithm develop an accurate world model and/or optimal policy while interacting with the world or a simulated environment over many episodes via a process of trial and error.
\end{enumerate}

Both approaches above create \emph{special-purpose} AI systems. The GPAI design approach departs from the above primarily by the fact that the resulting model can be integrated into systems that can perform a wide variety of tasks. We detail the steps below.

\begin{figure}[t]
    \centering
    \includegraphics[width=1.0\textwidth]{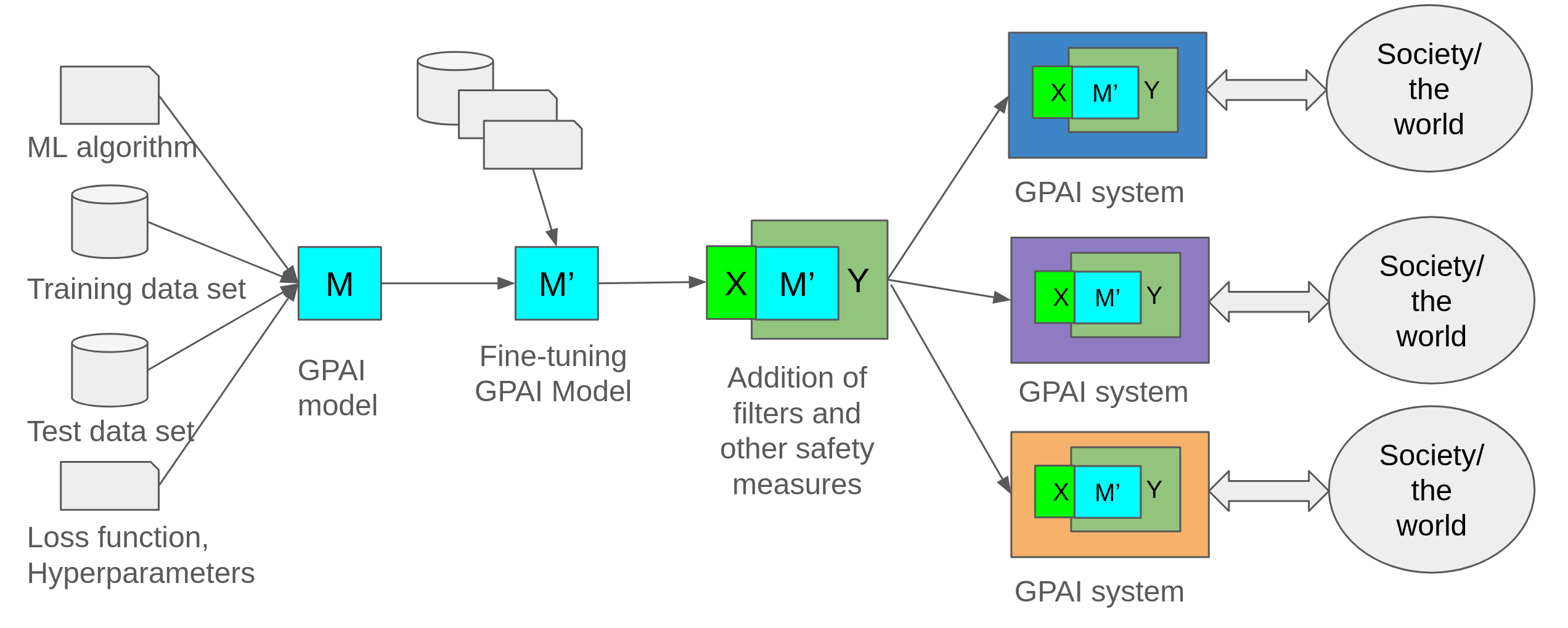}
    \caption{Example GPAI value chain: A GPAI model is created, modified, augmented by additional safety components, and then used to create several GPAI systems.}
    \label{fig:valuechain}
\end{figure}

Along the value chain in Figure \ref{fig:valuechain}, a GPAI model, typically an LLM, is first created by training a next-token predictor on a large corpus of text (or images, and potentially other media) obtained from the open internet and possibly several other sources. The resulting GPAI model $M$, however, may be a good next-token predictor with respect to the corpus, but the value chain foresees that it will eventually be configured and used for tasks very different from reliably predicting the next token based on the corpus.

Next in the value chain, a fine-tuning step is often used to change the model parameters, improving the model's ability to be used for a range of intended purposes. The parameters might be updated, for example, to make the model a more effective conversation engine, conducting conversations that are more useful, safe, desirable, or fair than those found in the training corpus. Parameters might also be updated to make the model more useful or reliable as a \textit{world model}, producing outputs that are more likely to be truthful statements about, or accurate depictions of, the real world.  

Parameter re-tuning is not the only way to make the model more useful or safe. The next step in the value chain may involve post-training enhancements by implementing various other capabilities or safety measures around the model, e.g., by creating a model API for interacting with the model that produces better results than direct interaction with the model itself.

Once the GPAI system is ready to be deployed, the GPAI system provider will have extensive means to configure or otherwise control the GPAI model so that it performs a more specific AI task. Some GPAI system providers may also design systems that remain “general,” where individual users are able, typically via a dialogue-based interface, to make the GPAI model perform a wide range of tasks. 

The deployed GPAI system or GPAI model is a dual-use technology\footnotemark, which will need to be regulated as such. The EU AI Act \cite{aiact}, for example, considers the possibility that bad actors could use highly powerful GPAI models to create “systemic risk” for society as a whole.  Based on these considerations, the Act places several value chain management requirements on the providers of GPAI models. 

\footnotetext{While “dual-use" usually refers to the fact that a given technology can be used for both civilian and military applications \cite{alic1994dual}, we remain flexible and use the term here to denote technology that can be used either for the benefit or harm of society.}

Another societal concern is that even “non-bad actors” along the value chain might, due to perverse market incentives, perform inadequate safety engineering. In response, the EU AI Act \cite{aiact} explicitly aims to “improve the internal market” of the Union by defining certain minimum requirements for safety engineering processes at key points along the GPAI value chain, similar to the way safety requirements exist in many other fields.

Systems of classification are often used to impose different sets of minimum requirements.  For example, with regards to the EU AI Act \cite{aiact}, GPAI models and systems that do not affect the Union market at all, while also not affecting any natural persons located in the Union, are excluded. In addition, a strong set of minimum safety engineering requirements for GPAI model providers applies only if their models are classified as having “systemic risk,” and GPAI system providers are subject to strong minimum safety engineering requirements only if their GPAI systems are classified as “high-risk AI systems.”  Market parties may, of course, choose to differentiate themselves from others by exceeding these minimum requirements.

\section{Risks and Limitations in Benchmarking}
\label{sec:benchmarking}
It is important to ensure that the chosen benchmark remains relevant: it should not be outdated, its performance should not be oversaturated (i.e., models should not perform near-perfectly on it), and it should effectively measure the intended properties. Proper documentation of benchmark collection, curation, and usage is needed to identify and track potential limitations or biases in the evaluation process. Once established, benchmarks offer a relatively low-cost method for ongoing performance assessment and can serve as early indicators of potential issues, signaling when more resource intensive methods, such as red teaming or the development of new safety techniques, may be required. 

However, there is a notable lack of safety-oriented benchmarks, meaning that certain critical aspects remain untested. As a result, it is difficult to determine whether safety-related goals have been met. Another concern is “benchmark leakage," which occurs when an AI model is trained or fine-tuned using data that overlaps with the evaluation dataset, leading to artificially inflated performance and skewed results. 

Following a recent common sense benchmarking survey \cite{davis2023benchmarks}, one can identify the following purposes of benchmarking: 

\begin{enumerate}
  \item Comparing different systems at a particular time 
  \item Measuring progress over time
  \item Measuring comparative performance on different tasks
  \item Serving as a measure in absolute terms (the ideal scenario)
  \item Providing resources for training or fine tuning
  \item Highlighting overlooked or understudied problems
  \item Demonstrating limitations
  \item Offering vivid examples (e.g., the trophy and suitcase example in the Winograd schema 
challenge \cite{levesque2012winograd})
  \item Improving the process of benchmark design and development
\end{enumerate}

In particular, we want to highlight points 6 and 9. Benchmarks should play a key role in identifying and measuring model capabilities, especially when they are overlooked or understudied. It is difficult to detect or effectively measure a model property if one does not know what they are looking for, or what is the exact scope of the analysis. Therefore, the development of benchmarks should be an ongoing, iterative process that evolves alongside model scaling.

\section{Risks from Autonomous AI Systems}
\label{sec:autonomous}

GPAI models can be sufficient building blocks to construct autonomous AI systems, for example by putting the GPAI into an input-output loop with its environment. This is already happening with large language models over API and open-weights. Examples include LangChain\footnote{\href{https://www.langchain.com/langchain}{https://www.langchain.com/langchain}}, Auto-GPT\footnote{\href{https://agpt.co/}{https://agpt.co/}}, Sakana AI’s AI scientist\footnote{\href{https://sakana.ai/ai-scientist/}{https://sakana.ai/ai-scientist/}}, and Cognition.ai’s Devin\footnote{\href{https://www.cognition.ai/blog/introducing-devin}{https://www.cognition.ai/blog/introducing-devin}} coding agent. These autonomous agent frameworks are sometimes called “agent scaffold” and are often open source.  
 
The deployment of autonomous AI systems poses systemic risks from malfunctions, malicious use, and other sources. These risks are exacerbated by the potential to cause damage while being undetected for long periods of time, or through uncontrolled self-replication. 

Risks from autonomous systems increase with growing levels of autonomy, with systems with higher levels of autonomy posing a superset of risks of systems with lower levels of autonomy. For example, those include risks from misuse (which occur even in non-transparent systems), reward tampering and specification gaming (which occur in adaptable systems), deceptive behavior and persuasion leading to unauthorized actions (which occur in cooperative systems), and risks of the AI system bypassing a sandboxed environment and self-replication (which occur in autopoietic systems). For definitions of these terms, see Section \ref{sec:autonomous_def}.
 
Currently, we believe risk management measures for autonomous GPAIs are lacking, and more work should be devoted to develop such measures. This is especially true in the context of highly autonomous AI systems configured to affect the real world without being subject to intensive human supervision. We generally recommend that providers and users of GPAI models refrain from using a GPAI model to construct autonomous systems, except when they can verify that adequate supervision and control mechanisms are developed and put in place. Given the current situation, we recommend that GPAI providers providing a GPAI model to downstream users should explicitly recommend against, or even forbid, the construction of highly autonomous AIs by using their GPAI model.

\subsection{Definition and Levels of Autonomy}
\label{sec:autonomous_def}

We define autonomy in the context of AI systems as the ability of an AI system to perform long 
sequences of actions in complex environments without necessary human supervision or correction. Autonomy of AI systems is generally either defined as the absence of human oversight (e.g., in ISO/IEC 22989 \cite{isoAI}, which defines autonomy as “characteristic of a system that is capable of modifying its intended domain of use or goal without external intervention, control or oversight”), or as the ability to act in and adapt to a complex environment \cite{simmler2021taxonomy}.

We therefore define levels for the degree of supervision and levels of autonomy in terms of capability. To define different levels of autonomy as a capability of AI systems, we use a taxonomy developed 
in previous research \cite{simmler2021taxonomy}.

AIs are assessed along five dimensions: 

\begin{itemize}
    \item \emph{Transparency}: An autonomous AI system is transparent if the supervisor can observe how the AI determines its actions and outputs from its inputs. 
    \item \emph{Determinacy}: An autonomous AI is deterministic if it always produces the same output upon receiving the same input.
    \item \emph{Adaptability}: An adaptable AI is an autonomous AI that can use previous inputs to change its outputs in response to some inputs, i.e., the AI can learn during operation.
    \item \emph{Cooperation\footnote{Note: The original paper has a category called “openness”, which we have split into cooperation and 
self-modification.}}: An autonomous AI system is open if it can adjust its interactions with the environment with the aid of other agents, such as adding or removing input channels with cooperation from humans and other AI systems. 
    \item \emph{Self-modification and replication}: An autonomous AI system fulfills this property if it can self-replicate without human aid, or if it modifies its own components and structure.
\end{itemize}

This then yields six levels of autonomy for AI systems: 

\begin{enumerate}
    \item Deterministic system: An AI system that is determined, transparent, unadaptable, non-cooperative, and non-self-modifying. 
    \item Non-transparent system: An AI system that is determined, non-transparent, unadaptable, non-cooperative, and non-self-modifying.
    \item Indeterministic system: An AI system that is indeterministic, transparent, unadaptable, non-cooperative, and non-self-modifying.
    \item Adaptable system: An AI system that is indeterministic, transparent, adaptable, non-cooperative, and non-self-modifying. 
    \item Cooperative system: An AI system that is indeterministic, transparent, adaptable, cooperative, and non-self-modifying.
    \item Autopoietic system: An AI system that is indeterministic, transparent, adaptable, cooperative, and self-modifying or self-replicating.
\end{enumerate}

\subsection{Supervision of Autonomous AI Systems}

Supervision of autonomous AI systems can detect and prevent harms stemming from systemic risks posed by such systems. This can be classified based on the entity performing the supervision as well as the degree of supervision. Such supervision can be performed by different entities: 

\begin{enumerate}
    \item One or more humans (human-in-the-loop), which is used in existing autonomous systems, such as autonomous vehicles \cite{emami2024human}. They can be distinguished into human-in-the-loop, human-on-the-loop and human-out-of-the-loop \cite{methnani2021let}:

    \begin{itemize}
        \item Human-in-the-loop supervision means that a human influences most or all decisions of the system, and performs that role throughout the entire operating period of the AI system. 
        \item Human-on-the-loop supervision occurs if a human supervises the AI system, but is usually not required to intervene for the continued functionality of the system.
        \item Human-out-of-the-loop is the case if a human never or rarely supervises or monitors the actions of the system, and never or rarely intervenes in the system during its runtime.
    \end{itemize}
    
    \item Non-autonomous (or semi-autonomous), reliable, and weaker GPAI systems (e.g., as proposed in previous research \cite{greenblatt2023ai}), potentially supplemented by occasional human review. 
    
    \item Rule-based programs supervising the AI systems' actions, as used in existing generative AI applications\footnote{E.g., \href{https://learn.microsoft.com/en-us/azure/ai-studio/concepts/content-filtering}{https://learn.microsoft.com/en-us/azure/ai-studio/concepts/content-filtering}}.
\end{enumerate}

Supervision of autonomous AI systems can occur to different degrees, which are outlined in ISO/IEC 22989 \cite{isoAI}: 

\begin{enumerate}
    \item No automation: The operator fully controls the system.
    \item Assistance: The system assists an operator. 
    \item Partial automation: Some sub-functions of the system are fully automated while the system remains under the control of an external agent.
    \item Conditional automation: Sustained and specific performance by a system, with an external agent being ready to take over when necessary.
    \item High automation: The system performs parts of its mission without external intervention.
    \item Full automation: The system is capable of performing its entire mission without external intervention.
\end{enumerate}

The level of supervision is independent of the type of entity performing the supervision, for example an autonomous system can be automated with assistance from a non-autonomous rule-based program, and conditional automation with regards to human oversight in situations where the system encounters a failure state.

\section{Proposed Risk Taxonomy}
\label{sec:risk_taxonomy}

This section surveys several proposed taxonomies attempting to categorize AI risks and proposes an alternative taxonomy.

\subsection{Background}
 The risk items in Sections \ref{Model development} through \ref{Impacts} were categorized under their respective headings, such as model development, model evaluations, and GPAI failure modes. However, this categorization was done thematically, and the compilation of headings is not intended to serve as a comprehensive taxonomy. Nevertheless, we believe that properly designed risk taxonomies can be useful for conducting risk assessment and mitigation in a systematic way. In this section, we discuss our proposed risk taxonomies, which consist of a taxonomy of harms and a taxonomy of risk source dimensions, drawing inspiration from the existing literature.

Existing risk taxonomies in the literature cover different aspects of risk. For example, the AI Risk Repository \cite{slattery2024ai} uses a Causal Taxonomy as well as a Domain Taxonomy, where each risk is associated with certain causes and domains. However, the two taxonomies are not cleanly separated by cause and effect, as some categories under the Domain Taxonomy can also be regarded as causal factors (e.g., AI system safety, failures, and limitations). 

Our proposed taxonomies aim to more clearly distinguish between cause and effect by separating a taxonomy of harms from a taxonomy of risk source dimensions. While not intended to be comprehensive or detailed, they provide a framework for risk categorization from a top-down perspective.

\subsection{Taxonomy of Harms}

Our proposed taxonomy of harms intends to address only harms, without including the causal factors that lead to them, using the definition of ``harm" outlined in Section \ref{sec:risk_management_terms}.

\begin{figure}[t]
    \centering
    \includegraphics[width=1\textwidth]{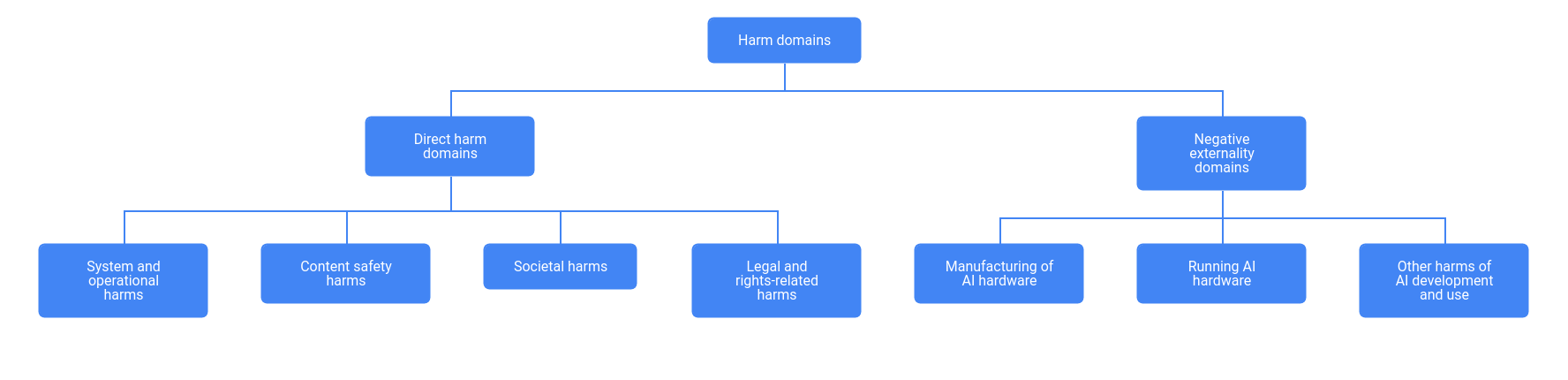}
    \caption{High-level categories of harms}
    \label{fig:taxonomy_harms}
\end{figure}

We propose the following taxonomy (Figure \ref{fig:taxonomy_harms}) for organizing harms into separate categories.  First, we divide them into two main ``harm domains."

\textbf{Harm domains:}
\begin{itemize}
    \item Direct harm domains
    \item Negative externality domains
\end{itemize} 

Direct harm domains cover harms directly resulting from the use of AI outputs or actions taken by AI agents, while negative externality domains cover indirect harms caused by the acquisition and use of resources required to run AIs, as well as other harms not directly caused by AI outputs.  
 
Their sub-domains are listed below, with examples of related harms listed under each sub-domain. 

\textbf{Direct harm domains:}
\begin{itemize}
    \item System and operational harms
        \begin{itemize}
            \item Security harms
                \begin{itemize}
                    \item Cybersecurity (e.g., cyberattacks)
                \end{itemize}
            
            \item Operational harms (e.g., automated decision-making)
                \begin{itemize}
                    \item Financial markets
                    \item Critical infrastructure
                    \item Other physical systems (e.g., transport)
                    \item Autonomous weapons
                \end{itemize}
        \end{itemize}

    \item Content safety harms
        \begin{itemize}
            \item Violence and extremism
            \item Hate and toxicity
            \item Sexual content
            \item Child harm
            \item Self-harm
            \item Dangerous content (e.g., CBRN)
        \end{itemize}
    
    \item Societal harms (e.g., misinformation, disinformation, and epistemic erosion)
        \begin{itemize}
            \item Political usage
            \item Economic harm
            \item Deception (e.g., fraud)
            \item Manipulation (e.g., deepfakes)
        \end{itemize}
    
    \item Legal and rights-related harms
        \begin{itemize}
            \item Discrimination and bias
            \item Privacy
            \item Criminal activities
        \end{itemize}
\end{itemize}

The direct harm domains consist of four main categories. For ``system and operational harms," the AI systems interact with other systems and industries, where a failure in an AI system could lead to failures of a wider scope. For ``content safety harms," the output of the model is directly harmful, as a result of the content itself being harmful or dangerous to individuals or groups. These are in contrast with ``societal harms," which are less direct but have more far-reaching effects on segments of society. Finally, ``legal and rights-related harms" concern either harms from illegal activities or harms from violations of human rights.

The direct harm domains and subdomains are largely adopted from AI Risk Categorization Decoded \cite{zeng2024ai} and NIST AI 600-1 AI RMF: GAI Profile \cite{nistGPAIprofile}. 

\textbf{Negative externality domains:}
\begin{itemize}
    \item Manufacturing of AI hardware
    \begin{itemize}
        \item Environmental harms from exploitation of natural resources
        \item Human rights harms from exploitation of human labor
    \end{itemize}
    \item Running AI hardware
    \begin{itemize}
        \item Environmental harms from energy usage
    \end{itemize}
    \item Other harms of AI development and use
    \begin{itemize}
        \item Societal inequality (individuals and companies who develop the best AIs get disproportionately powerful)
        \item Geopolitical harms (potential for conflict due to power imbalances)
    \end{itemize}
\end{itemize}

Direct externalities are mainly associated with manufacturing and running of AI hardware. Other externalities are less direct and involve more complex interactions, such as those caused by competitive dynamics of AI companies and countries.  

\subsection{Taxonomy of Risk Source Dimensions}
Existing taxonomies, such as the AI Risk Repository \cite{slattery2024ai}, use three dimensions for the Causal Taxonomy: Entity, Intent, and Timing. Kilian \cite{kilian2024beyond} categorizes risks into structural risks, accident / technical risks, and misuse, where there may be some overlap between these categories; while NIST \cite{nisttaxonomy} uses categories such as technical design attributes and socio-technical attributes.

Below, we propose several additional dimensions, with items within each dimension intended to be mutually exclusive and collectively exhaustive. The use of these dimensions offers additional perspective on how risk sources can be classified and, more importantly, facilitates the generation of associated risk management measures.

\begin{figure}[ht]
    \centering
    \includegraphics[width=1\textwidth]{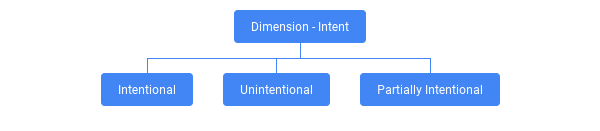}
    \caption{Types of intent}
    \label{fig:intent}
\end{figure}

\textbf{Dimension - Intent:} (Figure \ref{fig:intent})
\begin{itemize}
    \item Intentional
    \item Unintentional
    \item Partially intentional
\end{itemize}

Risks can be realized by intentional or unintentional actions, and in some cases the intent is difficult to establish. To manage these risks, rigorous evaluations and red teaming can be performed, guardrails can be put in place, and model release can be gradual, such that AI model malfunctions have either low likelihood or low probability of occurrence. To prevent intentional misuse, acceptable use policies can be in place, and for riskier models Know Your Customer (KYC) measures can also be implemented by model providers.

\begin{figure}[ht]
    \centering
    \includegraphics[width=1\textwidth]{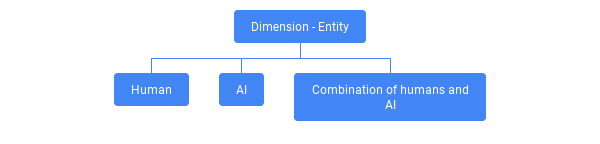}
    \caption{Types of entities}
    \label{fig:entity}
\end{figure}

\textbf{Dimension - Entity:} (Figure \ref{fig:entity})
\begin{itemize}
    \item Human
    \item AI
    \item Combination of humans and AI
\end{itemize}

A risk may be triggered by a human, where the AI serves merely as a tool, or by the AI acting autonomously with no human intervention, or it may involve a combination of both, with the human delegating some parts of decision-making to the AI. For risks where AI is the entity, these risks are exacerbated by an increase in the AI's level of autonomy. To manage risks involving AI as the trigger, appropriate levels of human oversight can be built-in. 

\begin{figure}[ht]
    \centering
    \includegraphics[width=1\textwidth]{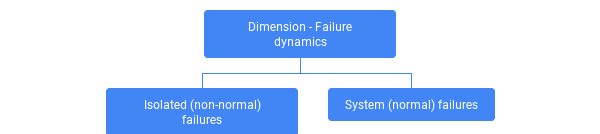}
    \caption{Types of failure dynamics}
    \label{fig:failure_dynamics}
\end{figure}

\textbf{Dimension - Failure dynamics:} (Figure \ref{fig:failure_dynamics})
\begin{itemize}
    \item Isolated (non-normal) failures
    \item System (normal) failures
\end{itemize}

In the context of Normal Accident Theory \cite{perrow2011normal}, normal accidents are those that ``could no longer be ascribed to isolated equipment malfunction, operator error, or acts of God.” We refer to these as ``system failures" (to be distinguished from ``systemic risks"), while the opposite would be ``isolated failures." For isolated failures, harms are consistent with the underlying failure modes. For example, an AI capable of producing false or misleading content would constitute risks related to misinformation and disinformation. Whereas for system failures, harms are not consistent with the underlying failure modes, or the harms are caused by interactions between different components within a system rather than from the failure of any specific component. One example of this is the 2010 Flash Crash, where a temporary stock market crash was partly caused by interactions between multiple trading algorithms, each of which was individually working as intended. Another example is the increasingly polarized society, potentially driven by algorithmic bias and amplification on social media applications, where recommendation algorithms simply recommended content baseed on the likelihood of user engagement, but the resulting harms were observed at a societal level. The nature of risks related to system failures makes them difficult to predict and manage in advance, but it underscores the importance of monitoring and identifying unexpected risks.

\begin{figure}[ht]
    \centering
    \includegraphics[width=1\textwidth]{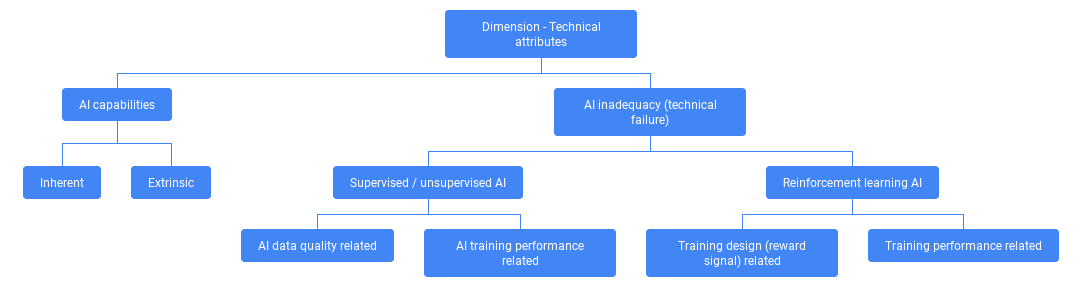}
    \caption{Types of technical attributes}
    \label{fig:technical_attr}
\end{figure}

\textbf{Dimension - Technical attributes:} (Figure \ref{fig:technical_attr})
\begin{itemize}
    \item AI capabilities
    \item AI inadequacy (technical failure)
\end{itemize}

An example of AI capabilities is that an AI might be capable of developing novel bioweapons. Whereas an example of AI inadequacy is a self-driving car causing an accident due to not being able to recognize certain objects. The boundary between capabilities and inadequacy is sometimes blurred. For example, when an AI generates falsehoods, it could be framed as either a capability of developing fiction, or an inadequacy in generating truthful content.

Under AI capabilities, there is another sub-dimension:

\textbf{Sub-dimension - Inherency of capabilities:}
\begin{itemize}
    \item Inherent
    \item Extrinsic
\end{itemize}

Inherent capabilities are inherent to the AI, whether they are deliberately trained or have emerged unintentionally. Extrinsic capabilities, on the other hand, are acquired through the use of external tools, such as LLM plugins. Inherent risks of AI capabilities are exacerbated through an increased amount of data and compute used for training, and an increase in modalities used as input or output for the AI (e.g., text, audio, video); while extrinsic risks involving AIs are exacerbated by an increase in the capabilities of the tools, and the interconnectivity between the AI and the available tools.  

Under AI inadequacy, there are sub-categories (which may not be exhaustive), depending on the type of AI:

\textbf{Sub-dimension - AI technical failure:}
\begin{itemize}
    
    \item For supervised or unsupervised AI
        \begin{itemize}
            \item AI data quality related
                \begin{itemize}
                    \item Biased training data
                \end{itemize}

            \item AI training performance related
                \begin{itemize}
                    \item Accuracy

            \item Reliability (related to precision)
            \item Robustness (related to model overfitting, spurious correlation, etc.)
                \begin{itemize}
                    \item Adversarial robustness
                \end{itemize}
            \end{itemize}
        \end{itemize}

    \item For reinforcement learning AI
        \begin{itemize}
            \item Training design (reward signal) related
                \begin{itemize}
                    \item Training that incentivizes unwanted behavior
                \end{itemize}
            
            \item Training performance related
                \begin{itemize}
                    \item Robustness (related to goal misgeneralization, reward hacking, etc.)
                \end{itemize}        
        \end{itemize}
        
\end{itemize}

As above, there are broadly two dimensions of technical failure modes: quality of data or input signal, and training performance. Due to a lack of transparency, it may be difficult to ascertain the type of technical failure that gives rise to a particular risk, and it is often a combination of several factors. Risks pertaining to AI failures are exacerbated by poor quality training data and imperfect training signals. Various measures can be implemented to improve the quality of the training data, and fine-tuning techniques can be used to disincentivize harmful model behavior.

\begin{figure}[ht]
    \centering
    \includegraphics[width=1\textwidth]{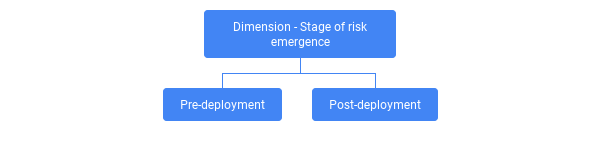}
    \caption{Stages of risk emergence}
    \label{fig:stage_risk}
\end{figure}

\textbf{Dimension - Stage of risk emergence:} (Figure \ref{fig:stage_risk})
\begin{itemize}
    \item Pre-deployment
    \item Post-deployment
\end{itemize}

For GPAIs or foundation models, risks emerge during training, prior to being repurposed and deployed in more specific AI systems or applications. Risk assessments can be conducted before deployment, and monitoring of AI models can occur as required throughout the deployment phase. In certain cases, version updates or model recalls may be warranted post-deployment.

\subsection{Insights on Categorizing Risk Sources into a Taxonomy}

A long list of risk sources can be structured into sections or categories in many different ways. Our main insight here is that there is no obvious best way to do it, as long as it is useful for effective implementation of risk assessment and management measures. We believe the completeness of the list is more important than its organization.

\end{document}